\documentclass[11pt,a4paper]{article}
\pdfoutput=1
\usepackage{jheppub}
\usepackage{amsfonts, amssymb, amsmath}
\usepackage{mathrsfs}
\usepackage[utf8]{inputenc}
\usepackage[russian,english]{babel}

\usepackage{color}
\definecolor{dark-gray}{gray}{0.20}
\definecolor{gray}{gray}{0.30}
\definecolor{light-gray}{gray}{0.80}
\definecolor{dark-red}{rgb}{0.7,0,0}
\definecolor{dark-green}{rgb}{0.1,0.4,0}
\definecolor{dark-blue}{rgb}{0.3,0.3,0.7}
\definecolor{light-blue}{rgb}{0.8,0.8,1}
\definecolor{blue}{rgb}{0,0,1}
\definecolor{red}{rgb}{1,0,0}
\definecolor{green}{rgb}{0,1,0}

\usepackage{hyperref}
\hypersetup{
	colorlinks=true,
	linkcolor=dark-blue,
	citecolor=dark-red,
	urlcolor=dark-blue,
	linktoc=page
}

\def\cA{{\cal A}}
\def\cB{{\cal B}}
\def\cD{{\cal D}}
\def\cF{{\cal F}}
\def\cI{{\cal I}}
\def\cK{{\cal K}}
\def\cL{{\cal L}}

\def\cN{{\cal N}}
\def\cO{{\cal O}}

\def\cV{{\cal V}}
\def\cW{{\cal W}}
\def\cJ{{\cal J}}
\def\cG{{\cal G}}
\def\cC{{\cal C}}
\def\cX{{\cal X}}
\def\cW{{\cal W}}
\def\cH{{\cal H}}

\def\U{{\rm U}}
\def\SU{{\rm SU}}

\def\OSp{{\rm OSp}}

\def\i{{\rm i}}

\newcommand{\be}{\begin{equation}}
\newcommand{\ee}{\end{equation}}
\newcommand{\bea}{\begin{eqnarray}}
\newcommand{\eea}{\end{eqnarray}}

\title{4d $\cN=2$ supergravity observables\\ from Nekrasov-like partition functions}

\author{Kiril Hristov}

\affiliation{Faculty of Physics, Sofia University, J. Bourchier Blvd. 5, 1164 Sofia, Bulgaria}
\affiliation{INRNE, Bulgarian Academy of Sciences, Tsarigradsko Chaussee 72, 1784 Sofia, Bulgaria}

\emailAdd{khristov@phys.uni-sofia.bg}

\abstract{
\noindent We reinterpret the OSV formula for the on-shell action/entropy function of asymptotically flat BPS black holes as a fixed point formula that is formally equivalent to a recent gluing proposal for asymptotically AdS$_4$ black holes. This prompts a conjecture that the complete perturbative answer for the most general gravitational building block of 4d $\cN=2$ supergravity at a single fixed point takes the form of a Nekrasov-like partition function with equivariant parameters related to the higher-derivative expansion of the prepotential. In turn this leads to a simple localization-like proposal for a set of supersymmetric partition functions in (UV completed) 4d $\cN=2$ supergravity theories. The conjecture is shown to be in agreement with a number of available results for different BPS backgrounds with both Minkowski and AdS asymptotics. In particular, it follows that the OSV formula comes from the unrefined limit of the general expression including only the so-called $\mathbb{W}$ tower of higher derivatives, while the on-shell action of pure (Euclidean) AdS$_4$ with round S$^3$ boundary comes from the NS limit that includes only the $\mathbb{T}$ tower. Backgrounds preserving less supersymmetry, such as the under-rotating black holes in flat space, the holographic squashed S$^3$, and the static/rotating twisted and non-twisted Kerr-Newman-like black holes in AdS$_4$ lead to a more general refined version of the corresponding gravitational blocks as dictated by the supersymmetric gluing rules. 
}
\date{\today}
\begin{document}
\maketitle


\section{Main conjecture}
\label{sec:intro}
The paradigm of holography, proposed in an explicit form by Maldacena in \cite{Maldacena:1997re} and further developed and tested  in innumerable papers, has shaped many of the ideas in high energy physics and string theory in the last decades. A closely related and much celebrated progress has been the successful microscopic counting of black hole degrees of freedom starting with the work of Strominger and Vafa \cite{Strominger:1996sh} and culminating in the Ooguri-Strominger-Vafa (OSV) conjecture \cite{Ooguri:2004zv}, made more precise by Denef and Moore \cite{Denef:2007vg}. These developments and many other related results present an overwhelming evidence that in many instances quantum gravity can be understood as a quantum field theory living in one dimension lower. A parallel major development in the last decades fueled by the seminal works of Seiberg and Witten \cite{Seiberg:1994rs,Seiberg:1994aj}, Nekrasov \cite{Nekrasov:2002qd}, and Pestun \cite{Pestun:2007rz}, has been the exact calculation of observables in supersymmetric quantum field theories. A major part of all the aforementioned progress has been achieved in 4d $\cN=2$ supersymmetric gravitational and field theories, which is precisely the case of interest here. 

Perhaps somewhat surprisingly, yet in no contradiction with the principles of holography, here we argue that much of the underlying structure of BPS observables in 4d $\cN=2$ supergravity is in fact {\it identical} to the structure of observables in 4d $\cN=2$ field theories. This general conclusion is based on a set of conjectural formulae argued to capture {\it all} supersymmetric partition functions of the theory that come from backgrounds with fixed points of their canonical isometry (see \cite{BenettiGenolini:2019jdz}), and as a limiting case a large set of backgrounds with fixed two-submanifolds. It is based on the combination of the result that the entropy function of black holes in AdS$_4$ can be written as a sum of {\it gravitational blocks} in two-derivative supergravity \cite{Hosseini:2019iad}, combined with the insights into higher derivative gauged supergravity solutions in \cite{Hristov:2016vbm} and \cite{Bobev:2020egg,Bobev:2021oku}, and inspired by the idea of \cite{Saraikin:2007jc} that extremal asymptotically flat non-BPS black holes lead to a refinement of the OSV formula to a full Nekrasov-like partition function. The remainder of this section is devoted to fully spelling out the conjecture, while the rest of the paper gives various arguments and explicit examples in support of it, without the pretense of giving exhaustive evidence. Part I of the conjecture proposes a higher derivative generalization of the gravitational building blocks of \cite{Hosseini:2019iad} making up the on-shell action of supersymmetric backgrounds, and is backed up most directly by a reinterpretation of the OSV formula that will be discussed in detail in section \ref{subsec:osv2}. Part II is instead an additional conjecture about the form of the full gravitational partition function on the same supersymetric backgrounds and is based on the Nekrasov conjecture \cite{Nekrasov:2003vi} and partially supported by the formalism of supergravity localization \cite{Dabholkar:2010uh,Dabholkar:2011ec,Dabholkar:2014wpa,Hristov:2018lod,deWit:2018dix}.

\subsection{Part I: the on-shell action}
\label{subsec:part1}
We consider matter-coupled 4d $\cN=2$ supergravity in the presence of $n_V$ vector multiplets\footnote{We assume additional hypermultiplets (or dual tensor multiplets) are not coupled to the other massless fields, either because of vanishing gauging or because they become massive. The case of interacting hypermultiplets deserves interest and would require a generalization of some of the formulae below.} and abelian electric gaugings (allowed to vanish). The Lagrangian of the two derivative theory, see e.g.\ \cite{Andrianopoli:1996cm}, is uniquely specified by the holomorphic prepotential $F_{2 \partial} (X^I)$, homogeneous of degree $2$ in terms of the symplectic sections $X^I$, $I=0, .., n_V$; and on the constant Fayet-Iliopoulos (FI) gauging parameters $g_I$. A number of possible higher-derivative (HD) invariants have been identified using the superconformal formalism, see e.g.\ \cite{Lauria:2020rhc}. We are going to argue that the full superspace integrals, or $D$-terms, vanish on the supersymmetric backgrounds we consider. We therefore consider only chiral superspace integrals, or $F$-terms; more precisely we look at two such terms coming from the square of the so-called Weyl multiplet \cite{Bergshoeff:1980is} and from the logarithm of the so-called kinetic multiplet \cite{Butter:2013lta,Butter:2014iwa}, see section \ref{sec:setup} for a careful discussion. We denote these HD invariants $\mathbb{W}$ and $\mathbb{T}$, respectively, and accordingly denote their lowest components (composite scalars) as $A_\mathbb{W}$ and $A_\mathbb{T}$.  The complete infinite-derivative supergravity Lagrangian is then fully specified by a polynomial expansion of the prepotential in the fields $A_\mathbb{W}, A_\mathbb{T}$, with holomorphic coefficients $F^{(m, n)} (X^I)$:
\be
\label{eq:1}
	F(X^I; A_\mathbb{W},  A_\mathbb{T}) := \sum_{m, n = 0}^\infty \, F^{(m,n)} (X^I)\, (A_\mathbb{W})^m\, (A_\mathbb{T})^n\ ,  
\ee
 where the leading piece is the two-derivative prepotential $F^{(0,0)} \equiv F_{2 \partial}$, and every coefficient $F^{(m,n)}$ is a homogeneous function of degree $2 (1-m-n)$ in $X^I$ and gives rise to terms with $2 (1+m+n)$ derivatives. Althought not strictly required by supersymmetry, string theory compactifications suggest that the Lagrangian defined by the above formula should be understood as a perturbative expansion in the gravitational coupling constant, $G_N^{(4)}$. While the explicit form of the higher derivative prepotentials $F^{(m,0)}$ is well-understood for compactifications on Calabi-Yau threefolds via the topological string, and one expects the refined topological string to relate to the full expansion of $F^{(m,n)}$ in flat space (see sections \ref{subsec:osv2} and \ref{subsec:under}), there is very little information about the prepotentials from AdS compactifications on Sasaki-Einstein manifolds. A more pragmatic approach, advocated in \cite{Bobev:2020egg,Bobev:2020zov,Bobev:2021oku}, is to use holography in order to fix these coefficients, which lead to explicit results for the four-derivative prepotentials $F^{(1,0)}$ and $F^{(0,1)}$ in a number of examples.

The theories defined by the full prepotential $F(X^I;A_\mathbb{W},  A_\mathbb{T})$ and the set of gauging parameters $g_I$ typically exhibit supersymmetric backgrounds asymptoting either to Minkowski space (e.g.\ when all $g_I = 0$) or to AdS$_4$, such that there exist rigorous ways to define and evaluate the finite on-shell action.\footnote{It is possible that the prepotential and gauging are incompatible with the condition for a maximally symmetric vacuum, see e.g.\ section 3.1 in \cite{Hristov:2018spe}. For the general form of the conjecture presented here we only need to assume a well-defined on-shell action.} As discussed at length in \cite{BenettiGenolini:2019jdz}, in order to do so it is useful to consider the canonical Killing vector field $\xi$ on a supersymmetric background $M_{4}$, defined as a bilinear of the Killing spinor $\epsilon$. For asymptotically AdS$_4$ solutions the on-shell action, $\cF (M_4)$, localizes on the fixed point set of $\xi$, which can consist of fixed points (nuts) and fixed two-dimensional submanifolds (bolts) \cite{BenettiGenolini:2019jdz}. Here we focus on the case of nuts (see later for comments on the bolts) and show that a resulting fixed point formula makes sense also for asymptotically flat black holes. Near such a fixed point, the Killing spinor becomes of definite chirality and the canonical vector is locally
\be
\label{eq:fixedpointvector}
	\xi = \varepsilon_1 \partial_{\varphi_1} + \varepsilon_2 \partial_{\varphi_2}\ , \qquad \varepsilon_2 / \varepsilon_1 \equiv \omega\ ,
\ee
with $\varphi_{1,2}$ being the standard polar angles on each copy of the two complex planes of the tangent space $\mathbb{C}^2$. Anticipating the results, we already defined the ratio of the two deformation parameters, $\omega$, which turns out to be the only physical and well-defined parameter \cite{BenettiGenolini:2019jdz}. The on-shell action is therefore dependent on the ratio $\omega$, as well as on what we call {\it Coulomb branch parameters} $\chi^I$ (for reasons apparent below), one complex parameter for each symplectic section $X^I$ (i.e.\ one extra Coulomb branch parameter than number of vector multiplets $n_V$).\footnote{This particular detail, which obscures somewhat the analogy with the field theoretic Coulomb branch parameters, is inherited from the off-shell formalism described in section \ref{sec:setup}.} We conjecture that the on-shell action, in the presence of arbitrary higher derivatives as defined by \eqref{eq:1}, of any supersymmetric background with fixed points of the canonical isometry can be written as a sum over contributions at each fixed point $\sigma$:
\bea
\label{eq:conj1}
\begin{split}
	\cF (M_4, \chi^I, \omega)  &=  \sum_{\sigma \in M_4} s_{(\sigma)}\, \cB (\kappa^{-1}\, X^I_{(\sigma)}, \omega_{(\sigma)})\ , \\
	\cB(X^I, \omega) &:= \frac{4 \i \pi^2\, F( X^I; (1-\omega)^2, (1+\omega)^2)}{\omega}\ ,
\end{split}
\eea
where $s_{(\sigma)} = \pm 1$ is aligned with the chirality of the Killing spinors at each $\sigma$,\footnote{We suppress the additional index $\pm$ that the authors of \cite{BenettiGenolini:2019jdz} assign to each fixed point, but still need to include the overall choice of sign. Since the chirality of the spinors of a single fixed point is conventional, we always choose the $+$ sign for the first fixed point of a given background.} and
\be
\kappa^2 := 8 \pi\, G_N^{(4)}\ ,
\ee
with $G_N^{(4)}$ the four-dimensional Newton constant. In the above formula the variables $X^I_{(\sigma)}$ are in general functions of $\omega$ and $\chi^I$ that need to be specified in dependence of the particular background and supersymmetry preservation. The identification of the $X^I_{(\sigma)}$, $s_{(\sigma)}$ and $\omega_{(\sigma)}$ at the different fixed points, together with the determination of an additional linear constraint $\lambda^{M_4} (g_I, \chi^I, \omega) = 0$,  are called the {\it gluing rules}. The constraint can be interpreted as restoring, in a supersymmetry-prescribed way, the correct number of Coulomb branch parameters. These rules need to be considered carefully for each specific BPS background, and we list four different gluing rules in table \ref{tab:1},\footnote{Note that at two derivatives one does not need to specify independently $s_{(\sigma)}$ as it can be reabsorbed in $\omega_{(\sigma)}$. The higher derivative corrections however depend on $\omega_{(\sigma)}$ in a more complicated way and it is important to identify separately the two quantities. This amounts to a minimal generalization of the {\it A} and {\it id} gluing rules of \cite{Hosseini:2019iad}.} describing all the explicit examples we discuss later. In general they contain information about the topology of the background $M_4$ and possible topological flavor charges, such as the magnetic charges $p^I$ of black hole solutions.
\begin{table}[ht]
	\begin{center}
		\setlength{\tabcolsep}{7pt}
		\renewcommand{\arraystretch}{1.3}
		\hspace*{-0.8cm}
		\begin{tabular}{|c || c | c | c | c |} \hline
	 gluing & {\it maximal (max)} & {\it identity (id)} & {\it A-twist (A)} & $S^3$ ({\it S}) \\  \hline
		 fixed points & $2$ & $2$ & $2$ & $1$ \\ \hline
		$\sigma = 1$ & $X^I_{(1)} = \chi^I - \omega p^I,$ & $X^I_{(1)} = \chi^I - \omega p^I,$ & $X^I_{(1)} = \chi^I - \omega p^I,$ & $X^I_{(1)} = (1+ \omega) \chi^I,$  \\ 
 &  $s_{(1)}=+, \omega_{(1)} = \omega$ &  $s_{(1)}=+, \omega_{(1)} = \omega$ &  $s_{(1)}=+,\omega_{(1)} = \omega$ & $s_{(1)}=+,\omega_{(1)} = \omega$ \\ \hline
	$\sigma = 2$ & $X^I_{(2)} = \chi^I + \omega p^I, $ & $X^I_{(2)} = \chi^I + \omega p^I, $ & $X^I_{(2)} = \chi^I + \omega p^I, $ & --- \\ 
 	& $s_{(2)}=-,\omega_{(2)} = \omega$ & $s_{(2)}=+,\omega_{(2)} = \omega$ & $s_{(2)}=+,\omega_{(2)} = -\omega$ & --- \\ \hline
 $\lambda (g_I, \chi^I, \omega)$ &  $ - \omega -1 $ & $g_I \chi^I - \omega -1 $ & $g_I \chi^I - 1$ & $g_I \chi^I - 1$ \\ \hline 
extra info &  $g_I= 0$ & $g_I p^I = 0$ & $g_I p^I = -1$ &  --- \\ \hline
		\end{tabular}
	\end{center}
	\caption{Four different gluing rules that can be applied to Eq.\ \eqref{eq:conj1}. The {\it identity} and {\it A-twist} rules were presented in \cite{Hosseini:2019iad} and match with analogous rules for gluing holomorphic blocks \cite{Beem:2012mb} for three-dimensional partition functions. The {\it max} rule corresponds to preserving maximal supersymmetry on S$^2$, related to static BPS black holes in ungauged supergravity. The {\it S}-rule is applied to asymptotically Euclidean AdS$_4$ backgrounds with squashed S$^3$ boundary (both $\U(1) \times \U(1)$ \cite{Plebanski:1976gy,Martelli:2011fu} and $\U(1) \times \SU(2)$ \cite{Martelli:2011fw} squashing) and is distinct from its three-dimensional counterpart in \cite{Beem:2012mb} due to the single fixed point rather than two.}
	\label{tab:1}
\end{table}

In many cases of interest, such as black holes, the variables $\chi^I$ and $\omega$ are the $U(1)$ chemical potentials conjugate to the conserved electric charges $q_I$ and angular momentum $\cJ$, respectively (while magnetic charges enter in the gluing rule for $X^I_{(\sigma)}$). In such cases one can also define the Legendre transform of the on-shell action, which is often called the {\it entropy function} \cite{Sen:2005wa}:
\be
\label{eq:entrop}
	\cI (M_4, \chi^I, \omega, q_I, \cJ) = - \cF (M_4, \chi^I, \omega) - \frac{8 \i \pi^2}{\kappa^2} (\chi^I q_I - \omega \cJ)\ ,
\ee 
which upon extremization reproduces the entropy of the black hole in question,
\be
\label{eq:entropy}
	S_\text{BH}  (M_4, q_I, \cJ) = \cI (M_4, \chi^I\Big|_\text{crit.}, \omega\Big|_\text{crit.}, \hat{q}_I, \hat{\cJ}) \in \mathbb{R}\ .
\ee
Note that the extremization is performed under the constraint $\lambda^{M_4} (g_I, \chi^I, \omega) = 0$, and the formula above automatically imposes an analogous constraint $\hat{\lambda}^{M_4} (g_I, q_I, \cJ) = 0$ on the set of conserved charges $\{q_I, \cJ \}$ by requiring that the entropy is a real quantity. Since the explicit expression for the resulting constraint $\hat{\lambda}^{M_4} (g_I, q_I, \cJ)$ has no obvious analytic form and is in general non-linear and strongly dependent on the explicit form of the prepotential, we resorted to the notation $\{\hat{q}_I, \hat{\cJ} \}$ above to denote that the constraint has already been imposed. Alternatively, one can view the imaginary part of the above equation as the definition of $\hat{\lambda}^{M_4} (g_I, q_I, \cJ)$,
\be
	\hat{\lambda}^{M_4} (g_I, q_I, \cJ) := {\rm Im} \left( \cI (M_4, \chi^I\Big|_\text{crit.}, \omega\Big|_\text{crit.}, q_I, \cJ) \right) = 0\ .
\ee

Let us pause here to observe that the so-called {\it gravitational building block} $\cB$, which is in complete agreement with the two-derivative answer in \cite{Hosseini:2019iad},\footnote{In order to have a manifest connection with the notation of \cite{BenettiGenolini:2019jdz}, we have rescaled the parameters $\chi^I, \omega$ in \cite{Hosseini:2019iad} as $\chi^I_\text{there} = 2 \chi^I_\text{here}, \omega_\text{there} = - 2 i\, \omega_\text{here}$. Also, due to the full expansion in $G_N^{(4)}$ (or $\kappa$), we had to include it inside the definition of the building block instead of leaving it as an overall normalization.}  now takes the form of (a logarithm of) a Nekrasov-like partition function where the expansion of the two higher derivative invariants is dictated by $A_\mathbb{W} \sim (\varepsilon_1 - \varepsilon_2)^2$ and $A_\mathbb{T} \sim (\epsilon_1 + \epsilon_2)^2$. The main upshot of the conjectural formula \eqref{eq:conj1} is that one only needs the gluing rules, fixed entirely by supersymmetry already at the two-derivative level, in order to evaluate the full higher derivative on-shell action.  Let us also define here the {\it gravitational Nekrasov partition function} as the exponent of the higher-derivative supergravity building block,
\be
\label{eq:Nek}
	Z^\text{sugra}_\text{Nek} (X^I, \omega) := e^{- \cB (\kappa^{-1}\, X^I,  \omega)} = \exp \left(- \frac{4 \i \pi^2\, F(\kappa^{-1}\, X^I; (1-\omega)^2, (1+\omega)^2)}{\omega} \right)\ ,
\ee
which will feature more prominently in part II of the conjecture.

Before giving explicit examples of BPS backgrounds and their corresponding gluing rules, we need to explain in which sense the on-shell action $\cF (M_4, \chi^I, \omega)$ depends on the continuous Coulomb branch parameters $\chi^I$ and deformation parameter $\omega$. We expect that there exists precisely one supersymmetric solution of (Euclidean) 4d $\cN=2$ supergravity that exhibits this on-shell action for each point on the moduli space spanned by $\chi^I$ and $\omega$,\footnote{In fact, due to the constraint we should treat $\omega$ in general as a complex quantity. The case of a single fixed point is an exception to this rule since there the geometric origin of $\omega$ is manifest and it cannot mix with the rest of the parameters.} under the specific gluing constraint $\lambda^{M_4} (g_I, \chi^I, \omega) = 0$, as advocated in \cite{Freedman:2013oja,Cabo-Bizet:2018ehj,Cassani:2019mms,Bobev:2020pjk}. For later purposes it is convenient to introduce the following novel notation that distinguishes between two basic types of supersymmetric backgrounds:
\begin{itemize}
	\item {\it special} (or {\it conformal}) solutions - we denote this way the solutions that {\it extremize} the function $\cI(M_4, \chi^I, \omega)$ in \eqref{eq:entrop} (or just $\cF (M_4, \chi^I, \omega)$ in \eqref{eq:conj1} in the lack of conserved charges conjugate to $\{\chi^I, \omega \}$), under the constraint $\lambda^{M_4} (g_I, \chi^I, \omega) = 0$. In other words these are the solutions obeying the so-called $\cI$-extremization \cite{Benini:2015eyy,Benini:2016rke} or $\cF$-extremization \cite{Jafferis:2010un,Jafferis:2011zi}, which indeed exhibit a symmetry enhancement. In the examples we consider $\cI$-extremization is obeyed by Lorentzian black holes exhibiting an AdS$_2$ symmetry in the near-horizon region, while $\cF$-extremization by pure (Euclidean) AdS$_4$ with round S$^3$ boundary. It is natural to think of the special solutions as defining the origin of moduli space, which is at the extremum\footnote{In fact there is no obvious reason why there should be a unique extremum, but this seems to be the case in the explicit examples we consider. The existence of more extrema is not excluded in the general formulae we present and does not lead to an obvious contradiction.} $\{\bar{\chi}^I (M_4), \bar{\omega} (M_4) \}$.

\item {\it Coulomb branch} solutions - these are all the remaining solutions, corresponding to the points on the moduli space $\{\chi^I, \omega \}$ away from the origin $\{\bar{\chi}^I (M_4), \bar{\omega} (M_4) \}$. As already stressed above, we expect that such a Coulomb branch solution exists for every point on the moduli space obeying the supersymmetric constraint. This has only been shown for a subset of the examples we consider but it is very natural to predict that the complete Coulomb branch can be explored. As the chosen notation suggests, we propose an interpretation in exact analogy with the moduli space of supersymmetric field theories. The Coulomb branch solutions are going to play an important role in part II of the conjecture.
\end{itemize}

We can now present in table \ref{tab:2} a short summary of the main solutions we consider as prime examples, in what type of (two-derivative) theories they exist and which gluing rules they obey. Given the discussion above, it is enough to only specify the {\it special} supersymmetric background we consider, knowing that each of them comes with its own space of Coulomb branch solutions.

\begin{table}[ht]
	\begin{center}
		\setlength{\tabcolsep}{7pt}
		\renewcommand{\arraystretch}{1.3}
		\begin{tabular}{|c | c | c | c | c |} \hline
	 special solution & gluing & $g_I$ & $F_{2 \partial}$ & constraint\\ \hline \hline
	static BH's in Mink$_4$ & {\it max} & $g_I = 0$ & $-\frac16\, c_{i j k} \frac{X^i X^j X^k}{X^0}$ & $\omega = -1$ \\ \hline
rotating$^*$ BH's in Mink$_4$ & {\it A} & $g_{i \neq 0} = 0$ & $-\frac16\, c_{i j k} \frac{X^i X^j X^k}{X^0}$ & $g_0 \chi^0 = 1$ \\ \hline \hline
round S$^3$ in AdS$_4$ & {\it S} & $g_I = 1$ & $-2 \i \sqrt{X^0 X^1 X^2 X^3}$ & $\sum_I \chi^I = 1$ \\ \hline
twisted BH's in AdS$_4$ & {\it A} & $g_I = 1$ & $-2 \i \sqrt{X^0 X^1 X^2 X^3}$ & $\sum_I \chi^I = 1$ \\ \hline
Kerr-Newman-AdS$_4$ & {\it id} & $g_I = 1$ & $-2 \i \sqrt{X^0 X^1 X^2 X^3}$ & $\sum_I \chi^I = (1+\omega)$ \\ \hline
		\end{tabular}
	\end{center}
	\caption{A summary of the explicit examples discussed in detail in sections \ref{sec:flat} and \ref{sec:AdS}. The suggested two-derivative prepotentials are typical examples of what string theory compactifications lead to. A clarification is needed for the second entry: this is the so-called under-rotating \cite{Bossard:2012xsa} or slow rotating \cite{Chow:2014cca} extremal {\it non-BPS} branch of black holes. The near-horizon geometry of these solutions is supersymmetric in the special case that $g_0 \neq 0$ \cite{Hristov:2012nu} and thus one can define a supersymmetric entropy function following the rules \eqref{eq:conj1}-\eqref{eq:entrop}. See section \ref{subsec:under} for more details.}
	\label{tab:2}
\end{table}

Finally, let us come back to the form of the gravitational building blocks $\cB$ that constitute the gravitational Nekrasov partition function \eqref{eq:Nek}. The two towers of higher derivative couplings are governed solely by the deformation parameter $\omega$, and a number of very important simplifications take place for different values of $\omega$ (or in the special case when $n_V=0$). One can then explicitly compare the table of special solutions and see which possibilities can be realized in which case. We distinguish the following five noteworthy limits:
\begin{itemize}
	\item {\it the unrefined limit}: $\omega = -1$ \\
		In this limit the on-shell action following from \eqref{eq:conj1} evaluates $A_\mathbb{T} = 0$, such that only the $\mathbb{W}$-tower of higher-derivative corrections can contribute, given by $F^{(m,0)}$ in \eqref{eq:1}. A check with table \ref{tab:2} tells us that the static BPS black holes in flat space sit precisely at the unrefined limit. This was expected since the OSV formula reproduces exactly the unrefined limit of the general conjecture, see section \ref{subsec:osv2}. Note that the unrefined limit is also reachable in the other black hole cases\footnote{In the holographic squashed S$^3$ case we find $\omega > 0$ and thus the unrefined limit is unattainable, see section \ref{subsec:sphere}.} but {\it not} for the respective special solutions. This means the unrefined limit corresponds to a point along the Coulomb branch and it is interesting to understand if these cases present a special interest.

	\item {\it the Nekrasov-Shatashvili (NS) limit}: $\omega = 1$ \\
		The NS limit is practically the opposite of the unrefined limit above, it corresponds to $A_\mathbb{W} = 0$ and therefore switching only the $\mathbb{T}$-invariant, $F^{(0,n)}$ in \eqref{eq:1}. This value of $\omega$ corresponds exactly to the case of the {\it round} holographic three-sphere and is therefore related to another special fully BPS solution, see section \ref{subsec:sphere}. Recently the authors of \cite{Binder:2021euo} indeed showed that the holographic round three-sphere case receives higher derivative corrections only from a certain type of $F$-terms, which here\footnote{In \cite{Binder:2021euo} the contributions are referred to as {\it chiral} $F$-terms from the point of view of the AdS$_4$ superalgebra. In the superconformal formalism {\it chirality} instead denotes superspace integration and {\it all} $F$-terms are by definition chiral. In the notation we use it is therefore more approapriate to use the distinction between $\mathbb{T}$ and $\mathbb{W}$ terms.} are identified as the $\mathbb{T}$-tower, in agreement with \cite{Bobev:2021oku}.

	\item {\it the Cardy limit}: $\omega \rightarrow 0$ \\
		The Cardy limit corresponds to the case when the on-shell action is actually divergent in the limit of vanishing $\omega$. This allows one to pick up only the leading term and is in fact applicable to the basic gravitational building block in \eqref{eq:conj1} itself. In the Cardy limit one therefore has access to the higher-derivative prepotential at the values $A_\mathbb{T} = A_\mathbb{W} = 1$. In the two-derivative case the Cardy limit of the gravitational building blocks matches precisely with the Cardy limit of the holomorphic blocks of the holographically dual 3d SCFT as shown in \cite{Hosseini:2019iad,Choi:2019dfu}. This limit is also applicable to the Kerr-Newman-like rotating black holes in AdS$_4$ with no twist, see section \ref{subsec:black holes no twist}, and the holographic squashed sphere in section \ref{subsec:sphere}.

	\item {\it the static/bolt limit}: $\omega = 0$ \\
		The cases when strictly vanishing $\omega$ is allowed can only exist in the special case when the gluing rules lead to a cancellation of the divergent terms. Since $\omega$ is conjugate to the angular momentum, this is the static limit of the under-rotating non-BPS black holes in flat space and the general rotating black holes with a twist in AdS$_4$, to be discussed in sections \ref{subsec:under} and \ref{subsec:black holes twist} respectively. At two derivatives the latter solutions are well-understood holographically via a dual field theory localization calculation, see \cite{Benini:2015eyy,Benini:2016rke,Zaffaroni:2019dhb} and references thereof. Going back to Eq.\ \eqref{eq:fixedpointvector} we notice that in this limit the fixed point actually blows up to a full two-dimensional submanifold. This is the so-called bolt case, which deserves a special attention since contributions from bolts need to generally be added to the conjectured formula \eqref{eq:conj1}. The existence of a smooth limit between nuts and bolts, emphasized in \cite{Hristov:2019xku,Hristov:2021zai}, allow us to also infer the contribution from a bolt, see Eq.\ \eqref{eq:boltconjecture} and the discussion around it. This is however {\it not} the most general possibility, since the neighborhood around the fixed two-manifold is in general a complex line bundle $\cO (-\mathfrak{p}) \rightarrow \Sigma_\mathfrak{g}$ with a Chern degree $\mathfrak{p}$, and the bolt limit we consider fixes $\mathfrak{p} = 0$.

	\item {\it the minimal/universal limit}: $n_V = 0$ \\
		The case of minimal supergravity with no additional vector multiplets allows for arbitrary $\omega$ but is nevertheless very special since one only has the single section $X^0$ and corresponding Coulomb branch parameter $\chi^0$. This limit is particularly important in all asymptotically AdS$_4$ examples since holographically it captures a {\it universal} sector of {\it all} dual three-dimensional field theories, as stressed in \cite{Azzurli:2017kxo,Bobev:2017uzs}. The minimal limit restricts immensely the freedom in the coefficients $F^{(m,n)}$ in \eqref{eq:1} since homogeneity in one variable means their only freedom is contained in an arbitrary multiplicative constant. This fact was used to obtain very general results for the on-shell action at four-derivatives in \cite{Bobev:2020egg,Bobev:2020zov,Bobev:2021oku}. The gluing rules in table \ref{tab:1} actually become degenerate in the minimal limit since the constraint makes sure the {\it identity} rule at a single point becomes equivalent to the {\it S}-rule, while the {\it A}-rule in minimal supergravity is instead relevant for the bolt limit. Another way of seeing the gluing universality is that there actually do not exist any Coulomb branch parameters anymore, except for the redundant one subjected to the constraint. One can then present a completely general form of the on-shell action at a given fixed point for all backgrounds, as shown in \cite{BenettiGenolini:2019jdz} at two-derivatives and \cite{Bobev:2020egg,Bobev:2021oku,Genolini:2021urf} at four-derivatives. The conjectural general formula \eqref{eq:conj1}, taken in the minimal limit, proposes a generalization to infinite derivatives.\footnote{Note however that results in \cite{Bobev:2020egg,Bobev:2021oku} that are specific to minimal four-derivative supergravity are from another point of view much more general than the conjectural gluing formula \eqref{eq:conj1}, since they hold for non-supersymmetric backgrounds as well.}

\end{itemize}

These five special limits are in fact the precise cases where our general formula \eqref{eq:conj1} can be tested most efficiently against available results. In this sense we have just listed the five pillars on which the general conjecture gets its support, in addition to the well-established two-derivative foundation in \cite{Hosseini:2019iad}.

\subsection{Part II: the partition function}
\label{subsec:part2}
To summarize the content of part I of the conjecture, we have presented a formula for the on-shell action of a large class of supersymmetric backgrounds in higher derivative supergravity. We should stress here that from an effective four-dimensional point of view we have just evaluated the on-shell action, which is an entirely classical quantity. We have so far neglected all perturbative and non-perturbative quantum corrections, both from loop diagrams of the four-dimensional dynamical fields and from external input from the UV completion that renders the quantum corrections finite.\footnote{Considered from a string theory point of view, the higher derivative terms in the four-dimensional effective action do come from a set of quantum (stringy) corrections to the original ten-dimensional two-derivative supergravity. Even from this perspective only a subset of quantum corrections are included, e.g.\ in the asymptotically flat compactifications the prepotential captures none of the $\alpha'$-corrections.} The fact that one can decompose the result in terms of the gravitational Nekrasov partition function \eqref{eq:Nek} however strongly suggests the structure of the quantum observables, i.e.\ the partition functions, whose classical limit has been evaluated by \eqref{eq:conj1}. We are therefore going to present a set of self-consistent formulae that incorporate this expectation, which naturally leads us to an {\it on-shell} localization-like formula in precise analogy to the Nekrasov conjecture \cite{Nekrasov:2003vi} for four-dimensional field theoretic observables decomposed into building blocks. The resulting formula closely resembles, and upon a particular identification is potentially coincidental, to the program of supergravity localization \cite{Dabholkar:2010uh,Dabholkar:2011ec,Dabholkar:2014wpa,Hristov:2018lod,deWit:2018dix}. Although we again specialize to the case of fixed points, it is clear that the proposal can be straightforwardly extended to the case of fixed two-submanifolds.  

In accordance with the above discussion, let us postulate the existence of a {\it quantum gravitational Nekrasov partition function} that depends on the same Coulomb branch parameters $\chi^I$ and deformation $\omega$,
\be
\label{eq:qgNek}
	Z^\text{q.g.}_\text{Nek} (X^I, \omega) := Z^\text{sugra}_\text{Nek} (X^I, \omega)\, Z^\text{UV}_\text{Nek} (X^I, \omega)\ ,
\ee
where the second factor contains {\it all} quantum corrections, perturbative or not, which are expected to assemble again in a form of a Nekrasov-like partition function.\footnote{Note that \cite{Nekrasov:2002qd} introduces a natural split of the partition function considered there in perturbative and instanton part, which has no obvious relation and should not be confused with the proposal here.} Since the effective four-dimensional supergravity is in general not a UV complete theory, the proposed $Z^\text{UV}_\text{Nek}$  must include the additional input from string theory outside the effective four-derivative action. Such external output is for example the full Kaluza-Klein (KK) tower of massive modes coming from the string theory, which generically cannot be decoupled at higher energies. Given the putative existence of such an object, we can use it to construct the (so far) {\it ad-hoc} function
\be
\label{eq:conj12}
	Z (M_4, \chi^I, \omega) := \prod_{\sigma \in M_4} Z^\text{q.g.}_\text{Nek} (X^I_{(\sigma)}, \omega_{(\sigma)})\ ,
\ee
supplemented with the set of gluing rules and a constraint $\lambda(g_I, \chi^I, \omega)$ that come from the classical analysis in the previous part of the conjecture. In order to simplify notation we implicitly included the overall signs $s_{(\sigma)}$ inherited from \eqref{eq:conj1} inside the product above.  This means that one can still use the information in tables \ref{tab:1} and \ref{tab:2} to build a complete answer for the set of considered special solutions and their corresponding Coulomb branch solutions away from the critical point. We propose that the above object is actually the {\it grand-canonical} gravitational partition function, which by construction evaluates to $\exp (-\cF (M_4, \chi^I, \omega) )$ in the classical limit. Notice that the implicit suggestion that the quantum gravitational building block is applicable to all supersymmetric observables is a somewhat misleading statement. One should always keep in mind that the UV completion of the effective 4d supergravity is strongly dependent on the asymptotic vaccuum, which in turn defines the particular string/M-theory embedding. The applicability of the quantum corrections to the basic building blocks is therefore necessarily specific to a particular UV completion. A similar cautionary note is valid already for the classical supergravity factorization, since the explicit realization of the prepotential \eqref{eq:1} is very distinct for Minkowski and AdS compactifications. 

It is equally natural to define the {\it microcanonical} gravitational partition function, which is the Laplace transform of the above expression with respect to the conjugate conserved charges,
\be
	Z (M_4, q_I, \cJ) := \int \left( \prod_{I = 0}^{n_V} {\rm d} \chi^I \right)\, {\rm d} \omega\, \delta(\lambda(g_I, \chi^I, \omega))\, e^{- \frac{8 \i \pi^2}{\kappa^2} (\chi^I q_I - \omega \cJ)}\, Z (M_4, \chi^I, \omega)\ ,
\ee
where we need to implement the constraint on the conserved charges $\hat{\lambda}^{M_4} (g_I, q_I, \cJ) = 0$, the microcanonical analog of the constraint $\lambda(g_I, \chi^I, \omega)$ (see the discussion below \eqref{eq:entropy}). The subtlety of needing a constrained Laplace transform is again dictated to us by supersymmetry, see also \cite{Benini:2016rke}.
We can alternatively rewrite this as
\be
\label{eq:conj2}
	Z (M_4, q_I, \cJ) = \int \left( \prod_{I = 0}^{n_V} {\rm d} \chi^I \right)\, {\rm d} \omega\, \delta(\lambda(g_I, \chi^I, \omega))\, e^{\cI (M_4, \chi^I, \omega, q_I, \cJ)}\, Z^\text{UV} (M_4, \chi^I, \omega)\ ,
\ee
with 
\be
\label{eq:corr}
	Z^\text{UV} (M_4, \chi^I, \omega) := \prod_{\sigma \in M_4} Z^\text{UV}_\text{Nek} (X^I_{(\sigma)}, \omega_{(\sigma)})\ .
\ee
It is always possible to perform the inverse transform,
\be
	Z (M_4, \chi^I, \omega) = \sum_{\hat{q}_I, \hat{\cJ}}\, e^{\frac{8 \i \pi^2}{\kappa^2} (\chi^I q_I - \omega \cJ)}\, Z (M_4, q_I, \cJ) \ ,
\ee
summing over the grid of appropriately quantized (in the UV complete theory) and {\it constrained} conserved charges obeying the implicit $\hat{\lambda}^{M_4} (g_I, q_I, \cJ) = 0$ derived from \eqref{eq:entropy}. This constrained sum is in turn replaced back by the constraint $\lambda^{M_4} (g_I, \chi^I, \omega) = 0$ valid in the grand-canonical ensemble.

The essence of the conjecture now is rather straightforward - we propose that the above expression of the microcanonical partition function is the quantum generalization of the entropy function.\footnote{Note that here we have ignored the possibility that other subleading saddles contribute to the gravitational path integral. There are several known examples in five dimensions where Euclidean solutions corresponding to quotients of the original solutions give a separate subleading contribution to the path integral, \cite{Murthy:2009dq,Aharony:2021zkr,Hong:2021bzg}. These solutions are {\it not} the same as the proposed Coulomb branch solutions and if existent in four dimensions need to be included separately. We do not discuss further this subtlety as it is tangential to the main ideas here and can be incorporated straightforwadly.} This means that in the black hole cases \eqref{eq:conj2} gives the {\it quantum entropy function} (QEF) \cite{Sen:2008vm} defined by Sen as the expectation value of a certain Wilson line operator in the near-horizon AdS$_2 \times$S$^2$ region.\footnote{In the case of the holographic S$^3$ background there are a number of exact dual partition function results in the literature. The original Airy function result for ABJM theory \cite{Aharony:2008ug} and other supersymmetric Chern-Simons theories \cite{Drukker:2011zy,Fuji:2011km,Marino:2011eh} was later generalized to include the squashing parameter \cite{Hatsuda:2016uqa} and the Coulomb branch parameters (interpreted holographically as the R-charge assignments) \cite{Nosaka:2015iiw}. It is not immediately clear whether these results should be compared with the grand-canonical or the microcanonical gravitational partition functions proposed here in \eqref{eq:conj12} and \eqref{eq:conj2}. We comment more on this in section \ref{subsec:sphere}.}  From the present point of view the purpose of the UV completion of 4d $\cN=2$ supergravity is to give a prescription for the evaluation of the missing piece, $Z^\text{UV}_\text{Nek} (X^I, \omega)$. It is important to observe that in the microcanonical partition function we essentially evaluate together the special solution with its full space of pertaining Coulomb branch solutions. The saddle-point evaluation of \eqref{eq:conj2} gives us back as a leading contribution exactly the special solution, but all the other Coulomb branch solutions contribute crucially with non-trivial subleading corrections. The finite-dimensional integral can therefore be interpreted as an {\it on-shell} localization formula, where every configuration described by the localization locus is actually a supersymmetric solution of (Euclidean) supergravity.

\subsubsection*{Relation to supergravity localization}
It is very interesting to observe that the form of the proposed microcanonical partition function is already in formal agreement with supergravity localization calculations for several of our examples of supersymmetric black holes \cite{Dabholkar:2010uh,Dabholkar:2011ec,Hristov:2018lod}, as well as for the holographic S$^3$ partition function \cite{Dabholkar:2014wpa}. The identification of these results with the proposed form in \eqref{eq:conj2} in fact leads to an additional physical understanding that is noteworthy. Let us focus for concreteness on the two examples of static supersymmetric black holes in Mink$_4$ and AdS$_4$ we mentioned so far, corresponding to the first and fourth row of table \ref{tab:2}, respectively (in the latter case we take directly in the static limit $\omega = 0$). In these cases there exist precise supergravity localization proposals for Sen's QEF \cite{Dabholkar:2010uh,Dabholkar:2011ec,Hristov:2018lod}, in agreement with the results of Denef and Moore \cite{Denef:2007vg}, where the localization locus defining the finite integration over parameters $\phi^I$ is interpreted as parametrizing the supersymmetric {\it off-shell} fluctuations around the near-horizon geometry. Written in the notation followed here, these localization calculations take the form
\be
\label{eq:Sen}
	Z_\text{QEF} (M_4, q_I) =  \int \left( \prod_{I = 0}^{n_V} {\rm d} \phi^I \right)\, \delta(\lambda(g_I, \phi^I))\, e^{\cI (M_4, \phi^I)}\, Z^\text{1-loop} (M_4, \phi^I)\, Z^\text{m} (M_4, \phi^I)\ ,
\ee
where the localization locus is spanned by the coordinates $\phi^I$. Additionally, there exists a clear prescription about the evaluation of the the 1-loop answer $Z^\text{1-loop} (M_4, \phi^I)$ which was shown \cite{Murthy:2015yfa,Hristov:2019xku} to exhibit the exact same factorization over fixed points as suggested in the putative Eq.\ \eqref{eq:corr}. Unfortunately there is not yet a clear prescription of how one should calculate the measure factor $Z^\text{m} (M_4, \phi^I)$, see e.g.\ \cite{Reys:2016ujf}, but the direct identification of \eqref{eq:corr} and \eqref{eq:Sen} suggests that it can also be split in building blocks. What is perhaps most surprising in such an identification is the conclusion that the off-shell localization locus in the supergravity localization formalism can be mapped to the on-shell space of Coulomb branch solutions via the simple $\phi^I = \chi^I$. Such an identification provides a natural extension of the supergravity localization program to the cases of black holes with rotation, for which the 1-loop contribution at each fixed point is already known as well \cite{Hristov:2021zai}. 

\subsubsection*{Relation to holography}
Let us also observe that the form of the grand-canonical and microcanonical partition functions proposed in \eqref{eq:conj12} and \eqref{eq:conj2} have a natural holographically dual interpretation, see e.g.\ \cite{Benini:2016rke} for an analogous discussion from the point of view of the topologically twisted index defined in \cite{Benini:2015noa,Benini:2016hjo}. The holographic interpretation of the proposed UV completed answer $Z (M_4, X^I, \omega)$ therefore translates into an all order expression at fixed $N$ of the holographically dual partition function, or index. A large class of supersymmetric three-dimensional partition functions are indeed known to admit an analogous factorization in terms of elementary building blocks, known as {\it holomorphic blocks} \cite{Pasquetti:2011fj,Beem:2012mb} (the K-theoretic lifts of vortex partition functions \cite{Dimofte:2010tz}) and/or {\it fibering operators} \cite{Closset:2018ghr}. It is thus tempting to speculate that the formal gravitational expressions for supersymmetric observables  in \eqref{eq:conj1} and \eqref{eq:conj2} might be useful in proving the AdS/CFT conjecture for the subset of supersymmetric observables discussed here. When making sense of our proposal holographically, it is very important that the gravitational Nekrasov partition function depends only on the ratio $\omega$ and not separately on $\varepsilon_{1,2}$ in \eqref{eq:fixedpointvector}. The three-dimensional building blocks naturally admit only one Omega-deformation parameter and it is therefore crucial that the four-dimensional analogs depend only on a single parameter as well, identified here precisely with the ratio of the two Omega-deformation parameters in four dimensions.\footnote{In this sense the chosen terminology of gravitational {\it Nekrasov} instead of {\it vortex} or {\it holomorphic} partition function is somewhat misleading, and we apologize to the reader for the potential confusion. Even if the gravitational answer exhibits only a single parameter $\omega$, it admits two opposite limits in the higher derivative expansion as discussed earlier, justifying the present choice.} It seems natural to expect that the full prepotential $F$ in \eqref{eq:1} relates in some precise way to the so called {\it twisted superpotential} $\cW$ of the dual three-dimensonal superconformal field theory placed on a circle, see \cite{Hosseini:2016tor,Closset:2017zgf,Closset:2018ghr}, evaluated on all possible Bethe vacua and not just the dominant one. It would be also very interesting to understand if the known structure of instantonic corrections of the S$^3$ partition function \cite{Drukker:2011zy,Fuji:2011km,Marino:2011eh,Hatsuda:2013oxa,Kallen:2013qla} can lead to a natural proposal for $Z^\text{UV}_\text{Nek} (X^I, \omega)$. Such bold attempts will however be pursued elsewhere. For the rest of this work we present the available supergravity calculations in support of the presented conjecture, only commenting in passing on the known holographically dual results where appropriate.

\subsection{Organization of the rest of the paper}
\label{subsec:outline}
Proving the complete conjecture, or even just part I of the conjecture in an example where the full refined version of the gravitational block is exhibited, is a rather ambitious goal that will be attempted elsewhere. Here we limit our evidence for the conjecture to an assorted summary of already established results in the available literature, which are going to agree with Eq.\ \eqref{eq:conj1} in one of the five special limits described above. We first discuss more carefully the superconformal formalism and the two types of $F$-term higher derivative corrections in section \ref{sec:setup}. We then split the discussion in two main parts, discussing separately the asymptotically flat and the asymptotically AdS$_4$ backgrounds in sections \ref{sec:flat} and \ref{sec:AdS}, respectively, which are kept independent of each other for the benefit of the readers that are more interested and familiar with one of the two cases. Since none of the results in sections \ref{sec:flat} and \ref{sec:AdS} is original, we adopt a minimalist approach and present only the main formulae, converted in a unified notation following section \ref{sec:setup}.

We can already give a more detailed account of the main references and their particular place in relation to the conjecture. In section \ref{subsec:osv2} we discuss the static single-centered and stationary multi-centered BPS black holes in flat space, which at two derivatives were analysed in \cite{Behrndt:1997ny} and \cite{Denef:2000nb}. Their higher derivative form was studied in \cite{LopesCardoso:1998tkj,LopesCardoso:1999cv,LopesCardoso:1999fsj,LopesCardoso:2000qm} and understood in a particularly suggestive form in \cite{Ooguri:2004zv}, where a conjecture about the corresponding partition function was put forward. In section \ref{subsec:under} we look at the extremal under-rotating asymptotically flat black holes, which at two derivatives can be understood by alternative descriptions both in ungauged supergravity where supersymmetry is preserved only asymptotically \cite{LopesCardoso:2007qid,Gimon:2007mh,Bena:2009ev,DallAgata:2010srl,Bossard:2012xsa}, and in the so-called {\it flat} gauged supergravity where supersymmetry is preserved only at the horizon via twisting \cite{Hristov:2012nu}. In the presence of higher derivatives these black holes were only analyzed in their static limit in \cite{Sahoo:2006rp,Banerjee:2016qvj,Hristov:2016vbm}, and their interpretation at the level of partition function was schematically discussed in \cite{Saraikin:2007jc}. In section \ref{subsec:universal} we describe the two derivative answer for on-shell action of all supersymmetric asymptotcally AdS$_4$ backgrounds \cite{BenettiGenolini:2019jdz}, and the four derivative generalization uncovered in \cite{Bobev:2020egg,Bobev:2021oku}. In section \ref{subsec:sphere} we focus on the holographic round S$^3$ background, which at two derivatives was understood in \cite{Freedman:2013oja} and was partially analyzed in \cite{Bobev:2021oku} at higher derivatives. The case of static \cite{Cacciatori:2009iz} and rotating \cite{Hristov:2018spe} asymptotically AdS$_4$ black holes with a twist is considred in section \ref{subsec:black holes twist}, whose entropy function was understood in \cite{Hosseini:2019iad}. Their corresponding Coulomb branch solutions in the static limit were discovered in \cite{Bobev:2020pjk}. The partial higher derivative description in the static limit was achieved in \cite{Hristov:2016vbm}, and a proposal for the corresponding partition function in the static limit was presented in \cite{Hristov:2018lod,Hristov:2019xku}. In the least well-understood and final example, we consider the Kerr-Newman-like black holes in AdS$_4$ with no twist in section \ref{subsec:black holes no twist}, discovered in \cite{Hristov:2019mqp} in the general matter-coupled two derivative theory. Their entropy function was understood in \cite{Hosseini:2019iad}, but the only available higher derivative analysis holds only in the  minimal supergravity limit \cite{Bobev:2020egg,Bobev:2020zov}.

\subsection{Open problems}
\label{subsec:part2}
From the discussion so far it should be clear that the main challenge left for future investigation is the proof, or potential disproof and/or correction, of the set of conjectural formulae presented above. The nature of part I of the conjecture is much more explicit and the path to proving the corresponding statements is conceptually clear, at least when asymptotically AdS$_4$ backgrounds are concerned. We should stress that we have completely neglected the important question of applying the procedure of holographic renormalization \cite{Emparan:1999pm,Skenderis:2002wp} to higher derivative off-shell supergravity, presenting a conjectural final result for the on-shell action that is secretly supposed to incorporate it. A rigorous derivation of this sort was achieved of \cite{BenettiGenolini:2019jdz} and \cite{Bobev:2021oku,Genolini:2021urf} in the minimal two derivative and four derivative cases, respectively, and one needs to generalize it to the full off-shell formalism including arbitrary number of vector multiplets. A very important step would be to understand the nature of the gluing rules concerning the Coulomb branch parameters $\chi^I$, and how precisely they arise on each background. At present we can only derive algorithmically the sign $s_{(\sigma)}$, correlated with the Killing spinor chirality, and the deformation $\omega_{(\sigma)}$ from the local form of the corresponding Killing vector. The set of proposed rules in table \ref{tab:1} is based on the detailed knowledge of the underlying backgrounds and their holographic duals rather than a general prescription, and should also be considered conjectural. 

An alternative and more practical goal, in the lack of such a general proof, would be to add further gluing examples, e.g.\ including the holographic Lens spaces \cite{Toldo:2017qsh}, the black spindles \cite{Ferrero:2020laf,Ferrero:2020twa} with gluing structure suggested in \cite{Hosseini:2021fge,Ferrero:2021ovq}, as well as more general punctures \cite{Bobev:2019ore} and possible multicentered solutions \cite{Monten:2021som}. It would also be interesting to understand properly the general case of fixed two-submanifolds with a nontrivial fibration that do not follow from a limit of the present results, i.e.\ the bolt examples in \cite{Toldo:2017qsh} and their generalizations with extra vectors. Such an accumulation of additional examples would perhaps make the general gluing principles more evident.

We should also stress that there are various other loopholes concerning off-shell supergravity and higher derivative invariants that we have largely ignored so far; we comment briefly on some of them in the coming sections. Perhaps the main point is that the form of the prepotential \eqref{eq:1} is itself possibly not exhaustive of all $F$-terms one might be able to add. Understanding this requires a careful examination and further classification of all higher derivative invariants. Additionally, we are going to argue on a case by case basis that $D$-terms do not contribute, but a general proof of this is also lacking.

There are also a couple of interesting extensions of the general results proposed here. The addition of gauged hypermultiplets will certainly influence the expression for the on-shell action, in the presence of non-compact gauged isometries we expect a Higgsing mechanism where some of the gauge fields and vector multiplet scalars become massive,  \cite{Hristov:2010eu,Hosseini:2017fjo,Hosseini:2020vgl}. In the formalism presented here, this might result either in additional constraints on the Coulomb branch parameters, or in a mechanism where the prepotential $F$ is replaced by an {\it effective} prepotential $F^*$. It is also natural to expect a very similar structure for supersymmetric observables in five-dimensional supergravity with $8$ real supercharges, where the precise 4d/5d off-shell map \cite{Banerjee:2011ts} suggests that the two derivative relation between building blocks in \cite{Hosseini:2019iad} can be extended to higher derivatives. The field theory analogy remains also intact since (the K-theoretic lift of) the Nekrasov partition function plays an important role in five dimensions \cite{Hosseini:2018uzp,Hosseini:2021mnn}, thus suggesting a five-dimensional version of the second part of the conjecture as well.

\section{Supergravity preliminaries}
\label{sec:setup}

In order to discuss rigorously the number of interesting examples related to the general conjecture, we need to introduce some important features of the formalism of conformal supergravity (or superconformal gravity), for comprehensive reviews see \cite{Mohaupt:2000mj,Lauria:2020rhc} and references therein, that is crucial for the proper analysis of higher derivative corrections. We are going to discuss the Lorentzian formalism here, with metric signature $(-, +, + ,+)$, since we look only at {\it special solutions} in the sense defined above, which consist of black hole spacetimes and pure AdS$_4$ space.\footnote{Strictly speaking, we need Euclidean signature for the choice of S$^3$ boundary, but practically we are only going to analyze the underlying AdS$_4$ vacuum.} In order to describe properly the complete set of Coulomb branch solutions one certainly needs to use the Euclidean formalism, developed in \cite{deWit:2017cle}, which is left for the future. We again stress that the rest of the this paper contains an assembly of already known results. Therefore our task in this section is to explain the general formalism and set-up unifying conventions, sparing the reader many details that are crucial in actually finding the supersymmetric solutions.

We should also caution that the superconformal formalism is an {\it off-shell} formulation of 4d $\cN=2$ supergravity with a number of additional unphysical symmetries and a corresponding set of {\it auxiliary} fields and multiplets, some of which do play an important part of the present analysis. Let us immediately give an example with the superconformal gravity multiplet, known as the Weyl multiplet, given by the following fields
\be
	\cW = \{e_\mu{}^a, \psi_\mu{}^i, b_\mu, A_\mu, \cV_\mu{}^i{}_j, T_{ab}^{ij}, \chi^i, D \}\ .
\ee
The only physical fields are actually the vielbein $e_\mu{}^a$ and the two gravitini $\psi_\mu{}^i$,  while the rest of the fields can either be directly gauge fixed away or can be solved for algebraically in terms of the other fields.\footnote{To be completely precise, the higher derivative corrections render some of the auxiliary fields physical, see \cite{Bobev:2021oku} for a discussion on the appearance an an additional massive gravity multiplet. One should therefore question the stability of the higher-derivative modes. Here we only look at supersymmetric backgrounds and assume a well-defined UV completion, which allows us to neglect such questions.} For later purposes we need to introduce the tensor field $T_{ab}^-:=T_{ab}^{ij} \varepsilon_{ij}$ and the real scalar $D$. The resulting actions and supersymmetry transformations therefore need to be subjected to a {\it gauge-fixing} procedure in order to arrive to the more commonly known {\it on-shell} Poincar\'{e} supergravity, see e.g.\ \cite{Andrianopoli:1996cm}. The gauge-fixing process is not unique and forces one to make certain choices that are in principle of no consequence for the physically observable quantities. However, since the Newton constant $G_N^{(4)}$ itself is an emergent physical parameter only after the gauge-fixing, it is important to keep track of how it is reinserted in the Lagrangian. In other words, the superconformal symmetry alone is not enough to tell us whether and how the higher-derivative corrections are suppressed in the coupling constant. This is why for the present purposes we need to make a choice consistent with what string theory and holography tell us. We comment more on this later.

Additionally, from the off-shell point of view, one already needs to start with the two-derivative Lagrangian of one auxiliary vector multiplet and one auxiliary hypermultiplet in order to arrive at the minimal on-shell formulation with the (on-shell) gravity multiplet only. As announced in the introduction, we consider non-minimal supergravity with a number $n_V$ of physical vector multiplets (denoted by $\cX^I$) in addition to the auxiliary one ($\cX^0$), and a single auxiliary hypermultiplet leading to the so-called FI gauging with the constant parametes $g_I$. The vector multiplets are given by the fields
\be
	\cX^I = \{ X^I, \Omega_i^I, W_\mu^I, Y^I_{ij} \},
\ee
where $W^I_\mu$ are the physical gauge fields, $\Omega_i^I$ the gaugini and $X^I$ are the already familiar complex scalars (one of which will be gauge-fixed). There is an additional triplet of auxiliary real scalars $Y^I_{ij}$ that will also play a role in what follows. 

We further consider two different four-derivative off-shell invariants, known as the Weyl$^2$ (here $\mathbb{W}$) and the T-log (here $\mathbb{T}$) \cite{Butter:2013lta,Butter:2014iwa}. If we want to allow for mixed terms between these two invariants, in agreement with the general form of \eqref{eq:1}, we need to define the composite field
\be
\label{eq:compositeA}
	\cA = c_1\, A_\mathbb{W} + c_2\, A_\mathbb{T}\ ,
\ee
which constitutes the lowest component of a composite chiral multiplet $\Phi$,
\be
	{\Phi}=({\cA},{\psi_i},{\cB_{ij}},{\cG^-_{ab}},{\Lambda_i},{\cC})\ ,
\ee
that we use to construct the theory,\footnote{Note that here we have rescaled by appropriate factors the definitions of both $A_\mathbb{W}$ and $A_\mathbb{T}$ in comparison with previous literature in order to minimize numerical factors entering \eqref{eq:conj1}. See below for the explicit values of these composite fields. In comparison to the standard notation in e.g.\  \cite{Banerjee:2016qvj}, we defined
\be
	A_\mathbb{W} = \frac{1}{64}\, A \Big|_{\cW^2}\ , \qquad A_\mathbb{T} = -\frac12\, A \Big|_{T (\log g_I X^I)}\ .
\ee} and $c_{1,2}$ are for now arbitrary dimensionless constants that parametrize a free mixing between the two invariants. When discussing the $\mathbb{T}$ in non-minimal supergravity, we have an extra choice for the chiral multiplet denoted $\Phi'$ in \cite{Butter:2013lta}, and which here we choose to be
\be
\label{eq:choicephiprime}
	\Phi' = \cX := g_I \cX^I\ .
\ee
Note that this choice is actually completely immaterial for our purposes here and was made to minimize notation (by using the convention of erasing the index for symplectic product between the FI parameters and the vector multiplets as above). Another standard choice is to pick $\cX^0$, to which we can for example stick in the ungauged case $g_I = 0$. The difference between any two arbitrary choices eventually leads to a higher-derivative invariant that can be obtained from a full superspace integral, i.e.\ it is a $D$-term \cite{deWit:2010za}, as shown in \cite{Butter:2014iwa}. We are going to argue that all $D$-terms must vanish on all supersymmetric backgrounds we consider and are therefore free to pick our $\Phi'$ without any consequence.

With the above choices, the composite field $\cA$ reads explicitly
\be
\label{eq:explicitA}
{\cA}= \frac{c_1}{64}\, (T^-_{ab})^2+c_2\,\Big(\frac{\Box_\mathrm{c} X}{\bar X}+\tfrac18\,\frac{\cF^{-}_{ab}\,T^{-\,ab}}{\bar X}-\frac{1}{8\,(\bar X)^2}\Big(Y^{ij}\,Y_{ij}-2\,\cF^{+}_{ab}\,\cF^{+\,ab}\Big)\Big)
\ee
where $\cF^{-}_{ab} := F^{-}_{ab}-\tfrac14\,\bar X\,T^-_{ab}$, as already hinted above $\{X, Y_{ij}, \cF_{ab} \} := \xi_I \{X^I, Y^I_{ij}, \cF^I_{ab} \}$, and $\pm$ superscripts here denoting (anti) self-duality. The above definition of $\cA$ uniquely determines the rest of the composite fields in the chiral multiplet $\Phi$, which appear explicitly in the Lagrangian.

The theory is then specified by the choice of prepotential $F(X^I; \cA)$, obeying the homogeneity property
\be
\label{eq:homogeneity}
	X^I\, F_I (X^I; \cA) + 2\, \cA\, F_{\cA} (X^I; \cA) = 2\, F (X^I; \cA)\ ,
\ee
where $F_I$ and $F_\cA$ denote the derivative of the prepotential with respect to $X^I$ and $\cA$.

In order to fix the conventions, let us present the bosonic part of the general {\it off-shell} action including an auxiliary vector and hypermultiplet, 
\begin{align}
\begin{split}
\label{eq:Lagrangian}
e^{-1} \, \cL =& \,\, -e^{-\cK}\, (\frac16\, R-D)
+ \Bigg[ {\rm i}\, \cD_\mu F_I \cD^\mu \bar X^I - \frac{\rm i}{8} F_{IJ} Y^I_{ij}
Y^{J, ij}  \\
&+\frac{\rm i}{4} F_{I J} \cF^{I, -}_{ab} \cF^{J, -, ab}
 +\frac{\rm i}{8} \bar F_I \cF^{I, -}_{ab} T^{-\,ab} + \frac{\rm i}{32} \bar F \, (T^-_{ab})^2
 + \frac{\rm i}{2} F_{\cA} { \cC} \\ &- \frac{\rm i}{8} F_{\cA
\cA} ({ \cB}_{ij} { \cB}^{ij} - 2 { \cG}^-_{ab}
{\cG}^{-ab}) 
+\frac{\rm i}{2} { \cG}^{-\,ab} F_{{\cA} I} 
\cF^{I, -}_{ab} -\frac{\rm i}{4} { \cB}_{ij} F_{{\cA} I} Y^{I, ij}
 + h.c. \Bigg] \\
&\, - \frac12 \varepsilon^{ij}\, \bar \Omega_{ab} \,
{\cal D}_\mu A_i{}^a \,{\cal D}^\mu  A_j{}^b
+\chi_H (\frac16 R+  \frac12 D) \\
&+\frac12\, G_{\bar a b}A^{i\bar a}\, g^2 \bar X X ( t^b{}_g\, t^g{}_d) A_i{}^d - \frac14\, A_i{}^a \bar\Omega_{a b}\,g Y^{ij}t^b{}_g A_j{}^g\ ,
\end{split}
\end{align}
with the function
\be
\label{eq:K-pot}
	e^{-\cK} = {\rm i} (\bar X^I F_I - X^I \bar F_I)\ ,
\ee
which formally resembles the K\"{a}hler potential but at this stage is {\it not} a function of the physical scalars, see below. Note that in these conventions the scalar curvature of AdS$_4$ is positive.

The last two rows above give the coupling of the auxiliary hypermultiplet that eventually leads to a scalar potential using the gauge fixing condition $A_i{}^\alpha = \chi_\text{H}^{1/2}\, \delta_i{}^\alpha$ and $t^i{}_j = \i\, \sigma^{3 i}{}_j$. See \cite{Hristov:2016vbm} for more details on the conventions we follow here and for careful discussion of how the FI gauging is introduced off-shell using the auxiliary hypermultiplet. Note one important change that involves both the notation and the convention for the FI terms, denoted $\xi_I$ in \cite{Hristov:2016vbm} and $g_I$ here, with the relation
\be
\label{eq:FIrescaling}
	\xi_I^\text{there} = \frac12\, g_I\ .
\ee
We should stress that this action contains terms with at most four derivatives if one considers the fields in the $\Phi$ multiplet as fundamental. Since they are composite fields that in turn depend on auxiliary fields, the final on-shell supergravity contains terms with two extra derivatives for each higher power of the $\cA$ appearing in the prepotential $F (X^I; \cA)$. The above form of the Lagrangian makes a number of choices in conventions that are not always followed in literature and deserve further explanation. We introduced a gauge coupling constant $g$ for the sole purpose of keeping the FI terms $g_I$ dimensionless, while instead refraining from the commonly introduced factor of $8 \pi$ on the left-hand side. We then propose that the physically meaningful way of restoring the gravitational coupling constant is by the following overall rescaling of the vector multiplet fields,
\be
\label{eq:rescaling}
	X^I \rightarrow \kappa^{-1}\, X^I\ , \quad W^I_\mu \rightarrow \kappa^{-1}\, W^I_\mu\ , \quad Y^I_{ij} \rightarrow \kappa^{-1}\, Y^I_{ij}\ ,
\ee
together with the choice $g = \kappa$. Via the equation of motion for the auxiliary field $D$, this choice introduces the correct factors of $\kappa$ also in the hypermultiplet sector and leads to the standard two-derivative on-shell Lagrangian upon choosing the prepotential $F(X^I; \cA) = F_{2 \partial} (X^I)$. Due to the homogeneity properties of the prepotential, the above rescaling means that we have introduced a manifest suppression of every additional power of $\cA$ in the prepotential (and Lagrangian) by a factor of $\kappa^2 = 8 \pi\, G_N^{(4)}$. One is of course always free to introduce additional meaningful (dimensionless) coupling constants via the choice of the coefficients $c_1$ and $c_2$ in \eqref{eq:compositeA} or directly in the coefficients $F^{(m,n)}$ in \eqref{eq:1}. Note that the above choice also means none of the fields in the Weyl multiplet carry any factors of $\kappa$, and it is easy to see that this also implies a lack of scaling for the composite multiplet $\Phi$. From now on we will directly work with the fields on the right hand side of \eqref{eq:rescaling} and not concern ourselves further with the factors of $\kappa$. Note that the above choice was already incorporated in the form of the building blocks in \eqref{eq:conj1}.

At this point we should emphasize that we have in fact {\it not} performed any part of the gauge-fixing procedure, but have merely rescaled the vector multiplet fields in order to introduce correctly the Newton constant in the theory. The full procedure of gauge fixing is explained carefully in e.g.\ \cite{Mohaupt:2000mj} and will not be needed here since we are going to present the supersymmetric solutions already in the off-shell formalism. Doing this leads to a major simplification - the off-shell supersymmetry variations (which can be explicitly found in the cited reviews) are actually {\it independent} of the specific Lagrangian. In fact the general supersymmetry variations include both the $Q$ and the $S$ transformation, but it is possible to take particular combinations of them in order to eliminate completely the $S$-variantions without the need of gauge-fixing. In the case of FI gauged supergravity, this was carefully discussed in sections 2 and 3 of \cite{Hristov:2016vbm}, which we follow completely up to the rescaling of the FI parameters in \eqref{eq:FIrescaling}. Ultimately, this is the origin of the statement that the gluing rules in table \ref{tab:1} hold in the exact same form for the general higher-derivative building blocks independently of the choice of prepotential. 

We should however explain more carefully the relation between the $(n_V+1)$ off-shell sections $X^I$ (after the rescaling in \eqref{eq:rescaling}) and the $n_V$ physical scalars denoted standardly by $z^I$. The rescaling above brings the correct powers of the Newton constant in the Lagrangian but does not yet lead to a canonical normalization of the Einstein-Hilbert term. In order to achieve this in a covariant way, by an abuse of notation introduced originally in \cite{Behrndt:1996jn}, we should acknowledge explicitly that the sections depend on the scalars $z^I$ and write
\be
\label{eq:offtoon}
	X^I = e^{\cK_{2 \partial} (z, \bar{z})/2}\, X^I (z)\ ,
\ee
where $\cK_{2 \partial} (z, \bar{z})$ is defined as in \eqref{eq:K-pot} based only on the two-derivative part of the prepotential and the variables $X^I (z)$, considered as projective coordinates that are implicitly already fixed in terms of the $n_V$ scalars $z^I$.\footnote{It is also common that the $X^I$ are denoted $L^I$, while the $X^I (z)$ are denoted by $Y^I$.} This overall rescaling leads automatically to the correct normalization of the Einstein-Hilbert term in the two-derivative theory. Furthermore, the physical scalars are standarly chosen as
\be
	z^I := \frac{X^I}{X^0} = \frac{X^I (z)}{X^0 (z)}\ ,
\ee
such that the overall scaling drops out and the abuse of notation introduced above is not misleading. The sections $X^I (z)$ are therefore the quantities that parametrize the scalar manifold in the {\it on-shell} theory, and which should be distinguished with their off-shell version via \eqref{eq:offtoon}.

It is clear from the above formulae, as well as the homogeneity properties of $F$, that the projective coordinates are still only defined modulo a multiplication by an arbitrary holomorphic function, which acts as a K\"{a}hler transformation on $\cK (z, \bar{z})$ . We are going to resort to a partial gauge-fixing at the expense of losing this remaining freedom, using a particular choice for both the D-gauge and the A-gauge (see again \cite{Mohaupt:2000mj}) to fix\footnote{In the ungauged case when $g_I = 0\, ,\, \forall\, I$, we cannot use this form and for the moment do not need to commit to a particular choice. An equivalent choice will be made in the process of discussing the particular example in section \ref{subsec:osv2}.}
\be
\label{eq:partialgauge}
	g_I X^I (z) = g_I \bar{X}^I (\bar{z}) = 1\ , \quad \Rightarrow \quad X (z) = \bar{X} (\bar{z}) = 1\ .
\ee
This partially gauge-fixed variables allow us to relate explicitly to the available two-derivative results usually written in terms of $X^I (z)$, while in the same time still keeping a largely off-shell point of view.
 
With the gauge choice above, and assuming that the scalars are constant,\footnote{Note that we do not need to consider backgrounds with constant scalars below. This simplification takes place automatically when zooming in on the fixed points of the canonical isometry.} we can simplify the expression for $A_\mathbb{T}$,
\be
\label{eq:finalAWAT}
	A_\mathbb{W} = \frac{(T_{ab}^-)^2}{64}\ , \, A_\mathbb{T} = \frac16\, R - D +\frac{e^{-\cK_{2 \partial} (z, \bar{z})}}{8}\Big(2\,\cF^{+}_{ab}\,\cF^{+\,ab}+e^{\cK_{2 \partial} (z, \bar{z})/2}\, \cF^{-}_{ab}\,T^{-\,ab}-Y^{ij}\,Y_{ij}\Big)\ ,
\ee
where we used explicitly the definition of $\Box_\mathrm{c}$ in \eqref{eq:explicitA}. These expressions will be explicitly evaluated on the specific backgrounds we consider and seen to be in agreement with the proposed evaluation in \eqref{eq:conj1}.

Finally, we can comment in more detail on the possible $F$-terms that we consider. One important point to notice is that the presented Lagrangian is indeed uniquely specified by a prepotential with the formal expansion as in \eqref{eq:1}, but due to \eqref{eq:compositeA} the two supersymmetric invariant considered here can be put together in the function $F(X^I; \cA)$. Consequently, the full expansion in \eqref{eq:1} is actually constrained since it can be written as an expansion in $\cA$ instead of $A_\mathbb{W}$ and $A_\mathbb{T}$, e.g.\ $F^{(1,0)} (X^I) \propto F^{(0,1)} (X^I)$ in \eqref{eq:1}, etc. This form was forced on us by requiring the two invariants to mix with each other. We could alternatively choose to keep them unrelated and obtain two independent towers with $F^{(m,0)} (X^I)$ and $F^{(0,n)} (X^I)$ instead. We should however stress that we have only added only two possible invariants, which are in particular known to contribute at four derivatives.\footnote{In fact already at four derivatives there has been yet another proposed invariant, \cite{deWit:2006gn,Kuzenko:2015jxa,Hegde:2019ioy}, which has rather different properties and makes use of an auxiliary tensor multiplet instead of a hypermultiplet. We conjecture that such type of invariants do not appear from string theory compactifications as partially argued in \cite{Bobev:2021oku}, but further analysis is needed for a definiteve statement.} It is expected that further $F$-term invariants come at six and higher number of derivatives. In view of the structurally pleasing form of the proposed gravitational building blocks in \eqref{eq:conj1}, we expect that the higher-order $F$-terms actually conspire to complete the full expansion in \eqref{eq:1} rather than appear as separate contributions that spoil the symmetry of the expressions. It is also worth remarking that the explicit, but rather incomplete, knowledge of string theory compactifications does not seem to require at present any generalization to the Lagrangian proposed here based on \eqref{eq:compositeA}. We remark further on this point when discussing explicitly the asymptotically flat and the asymptotically AdS theories below.

\section{Asymptotically flat solutions}
\label{sec:flat}
Here we focus on two special classes of asymptotically flat black hole solutions with supersymmetry preserved (at least) near the event horizon. The general form of two-derivative prepotential given in table \ref{tab:2},
\be
\label{eq:usualcubicprepotential}
	F_{2 \partial} = -\frac16\, c_{i j k}\, \frac{X^i X^j X^k}{X^0}\ ,
\ee
corresponds to the models with a known string theory origin from Calabi-Yau compactifications, where the coefficients $c_{i j k}$ have a geometric interpretation as the intersection numbers of the non-trivial cycles on the underlying manifold. Note that only a subset of the possible coefficients $c_{ijk}$, obeying an additional identity, give rise to scalar manifold that is a symmetric space. Such a condition is not strictly necessary, but technically leads to major simplification in the description of the black hole solutions due to the existing quartic-invariant formalism, see e.g.\ \cite{Hristov:2018spe} and references therein.

\subsection{BPS black holes}
\label{subsec:osv2}
We first discuss the prototypical example of supersymmetric black holes in ungauged supergravity that preserve half of the supersymmetries. As a limit of the general formalism described above, ungauged supergravity is simply reached via
\be
	g_I = 0\ , \qquad \forall I\ .
\ee
The asymptotic vacuum of these solutions is the maximally supersymmetric Minkowski space, while zooming in on the near-horizon geometry of the black holes leads to another fully-BPS complete solution - the AdS$_2 \times$S$^2$ geometry with equal radii, known as the Bertotti-Robinson spacetime. Note that due to supersymmetry, we are not actually forced to describe a single black hole, but an arbitrary collection of black hole centres staying at equilibrium with each other. The single-center solution is rotationally invariant and therefore static, while the multi-centered solutions allow for a relative rotation between the centres as long as the overall conserved angular momentum measured at infinity remains zero. The full black hole solutions are governed by the so called {\it attractor mechanism} \cite{Ferrara:1995ih,Ferrara:1996dd,Ferrara:1996um,Behrndt:1997ny}, where the scalars flow from their arbitrary asymptotic values to a fixed locus at each horizon in dependence on the corresponding conserved electro-magnetic charges. 

In the present work we are interested in fixed points of the canonical isometry, which are situated precisely on the corresponding horizons of each centre. We will therefore focus without any loss of generality directly on a single instance of the Bertotti-Robinson geometry and look at its properties first in the two derivative theory and then in the presence of higher derivatives. Note that the near-horizon analysis is enough to exhibit the attractor mechanism and to show how the single-centered black holes' entropy function fits in the proposed form of the general building blocks in \eqref{eq:conj1}. The multi-centered solutions should naturally obey the same rules since the sum over different fixed points is automatically built in. We are however going to neglect here more subtle points about the domains of existence of different solutions and the possible decay of multi-centered configurations, known as walls of marginal stability \cite{Denef:2000nb}.

Before moving to discuss in more detail the attractor mechanism, let us stress here that we are strictly looking at the {\it special solutions} here, corresponding to the well-known Lorentzian black holes. We are not aware of the existence of known Euclidean solutions corresponding to the same electromagnetic charges but keeping the value of the scalars arbitrary, which would provide the corresponding Coulomb branch solutions. It is however well-known that similar solutions exist in general and, as postulated in part I of the conjecture, we predict they can be found also in this case.

\subsubsection*{Two derivative solutions}
The fully BPS Bertotti-Robinson background preserves both the AdS$_2$ and the S$^2$ symmetries, packaged together in the full supersymmetry algebra $\SU(1,1|2)$. The Killing spinors analogously decompose into the standard two-dimensional spinors on the two spaces \cite{Lu:1998nu}, see e.g.\  \cite{Hristov:2012bk,Murthy:2015yfa}. To illustrate the main point, see explicitly App. B in \cite{Murthy:2015yfa}, where the explicit Killing spinors and their bilinears were spelled out in Euclidean signature. Using the standard coordinates $\tau, \eta$ on $\cH_2$ (Euclidean AdS$_2$) and $\theta, \varphi$ on S$^2$, we find the canonical isometry
\be
	\xi = - \partial_\tau + \partial_\varphi\ .
\ee
It is easy to see that the canonical isometry has precisely two fixed points, situated at the centre of AdS$_2$ and the south (SP) and north pole (NP) of the two-sphere, respectively. In the neighborhood of the two fixed points we can see that $\tau$ and $\varphi$ are precisely the polar angles of the two complex planes of the tangent space, c.f.\  \eqref{eq:fixedpointvector}. Moreover,one can explicitly check that the Killing spinors flip chirality when evaluated at the two poles, which gives rise in our conventions to
\be
	\omega_\text{SP} = \omega_\text{NP} = \omega = -1\ , \qquad s_\text{SP} = - s_\text{NP} = 1\ .
\ee
We therefore propose a new gluing rule, which we called {\it max}-gluing in table \ref{tab:1}, which results in the constraint $\omega = -1$ while flipping the overall sign $s$ for the two fixed points as above. The gluing rule for the Coulomb branch parameters is instead following precisely the two other rules on a two-sphere, the {\it id} and the {\it A} rules. We should however note that the fully BPS Bertotti-Robinson solution exists only when all gauging parameters vanish, and in this sense the {\it max}-gluing is much more restrictive than its half-BPS counterparts. Apart from the rule that reduces the constraint automatically to the value of $\omega$ found above, there are no further restrictions on the magnetic charges\footnote{We should emphasize here that supersymmetry indeed allows for an arbitrary set of electric and magnetic charges, which is also built in the gluing rule. However, the higher dimensional interpretation of the particular charge $p^0$ leads to additional complications since it corresponds to a fibration over the 5d circle and its presence undermines the five-dimensional black string interpretation. Therefore the explicit relation with the topological string partition function discussed later will require $p^0 = 0$.} $p^I$ and the Coulomb branch parameters $\chi^I$, again in precise agreement with the explicit background solution. 

Let us now focus on the attractor mechanism for the static BPS black holes considered here. The nature of the attractor mechanism and the resulting entropy function are well-known for their central role in the microscopic entropy counting via the Cardy formula, \cite{Strominger:1996sh,Maldacena:1997de}, and the influential work of \cite{Ooguri:2004zv} suggested that they follow precisely from a fixed point formula. Let us first write down the main attractor equations that fix completely the scalars in terms of the charges. In the notation of section \ref{sec:setup},
\be
\label{eq:attractor}
	\frac12\, \left( e^{\i \alpha} X^I (z) + e^{-\i \alpha} \bar{X}^I (\bar{z}) \right) = p^I\ , \qquad  \frac12\, \left( e^{\i \alpha} F_I (z) + e^{-\i \alpha} \bar{F}_I (\bar{z}) \right) = q_I\ ,
\ee
where the arbitrary constant phase $\alpha$ is correlated with phase of the Killing spinor and can be chosen freely without loss of generality. Based on these equations, \cite{Ooguri:2004zv} introduced the {\it real} variables $\phi^I$ via
\be
	e^{\i \alpha} X^I (z) = p^I + \frac{\i}{\pi}\, \phi^I\ .
\ee
In terms of these variables, the entropy function can be written in a mixed ensemble where $\phi^I$ are the chemical potentials, conjugate to the electric charges $q_I$,
\bea
\label{eq:OSVform}
	\begin{split}
		\cF_\text{OSV} (\phi^I, p^I) &= \frac{\i \pi}{2\, G_N^{(4)} } \left( F_{2\partial}(p^I + \frac{\i}{\pi}\, \phi^I) -  F_{2\partial}(p^I - \frac{\i}{\pi}\, \phi^I) \right)\ , \\
	\cI_\text{OSV} (\phi^I, p^I, q_I) &=  \cF_\text{OSV} (\phi^I, p^I)  + \frac{1}{G_N^{(4)} }\, \phi^I q_I\ ,
\end{split}
\eea
where we have introduced the appropriate factors of the Newton constant, set to unity in the original reference. The extremization of the entropy function $\cI_\text{OSV}$ leads back to the attractor equations above. This is the celebrated OSV formula in the two derivative case. We are going to discuss its generalization to higher derivatives in due course.

We can also evaluate the on-shell action and the entropy function that follow from the general formula \eqref{eq:conj1} applied to the two derivative prepotential using the {\it max}-gluing,.  
\bea
\label{eq:gluingform}
	\begin{split}
		\cF (\chi^I, p^I) &= - \frac{\i \pi}{2\, G_N^{(4)} } \left( F_{2\partial}(\chi^I + p^I) -  F_{2\partial}(\chi^I - p^I) \right)\ , \\
	\cI (\chi^I, p^I, q_I) &=  - \cF (\chi^I, p^I)  - \frac{\i \pi}{G_N^{(4)} }\, \chi^I q_I\ .
\end{split}
\eea
The two sets of formulae \eqref{eq:OSVform} and \eqref{eq:gluingform} agree precisely upon the the identification
\be
	\phi^I = - \i \pi\, \chi^I\ ,
\ee
and therefore the critical point for $\chi^I$ is purely imaginary and gives the black hole entropy via \eqref{eq:entropy}. Note that in this case it is particularly easy to infer the constraint on the conserved charges that follows from the constrained on the chemical potentials. $\omega = -1$. Upon the addition of the term $\i \omega \cJ$ to the entropy function above, we would simply discover that the constraint on the charges dictated by \eqref{eq:entropy} is just $\hat{\lambda} = \cJ = 0$, keeping the electric charges unconstrained, $\hat{q}_I = q_I$. This was of course guaranteed to happen since the class of black holes we are considering here is only static. We therefore jumped a step when presenting \eqref{eq:gluingform} and already implemented this rule. Furthermore, the background values of the sections (which are constant on the sphere and therefore equal at the two poles, $X^I_\text{SP} (z) = X^I_\text{NP} (z)$) relate to the critical values of the gluing parameters $X^I_{(1),(2)}$,
\be
\label{eq:identificationosv}
	X^I_\text{SP} (z) = e^{-\i \alpha}\, X^I_{(1)} \Big|_\text{crit.}\ , \qquad (X^I_\text{NP} (z))^* = -e^{\i \alpha}\, X^I_{(2)} \Big|_\text{crit.}\ ,
\ee
similar to the analogous formulae for the {\it id}-gluing example in section \ref{subsec:black holes no twist}.

Let us finally point out that the OSV formula was obtained by a smart rewriting of the on-shell answer for the black hole entropy, and therefore it was not proposed to hold for arbitrary values of the parameters $\phi^I$, which were defined from the start to be real. On the other hand, the proposed on-shell action based on the parameters $\chi^I$ using the {\it max}-gluing formally seems identical using the relation \eqref{eq:identificationosv}. Upon extremization of the entropy function this identitiy is automatically imposed and we find that $\chi^I\Big|_\text{crit.}$ are purely imaginary. However, the parameters $\chi^I$ are not a priori fixed to be imaginary, and as suggested in part I of the conjecture there should be Coulomb branch solutions that explore the full parameter space for $\chi^I$. In the lack of their explicit knowledge the distinction at present remains a mere technicality, but future analysis could render it more important.

\subsubsection*{Higher derivatives and the unrefined limit}

The Bertotti-Robinson background, due to its high amount of symmetry and historically leading significance for black hole microstate counting. is the most comprehensively studied and well-understood solution in off-shell supergravity. The Weyl$^2$, or $\mathbb{W}$ invariant was considered in a series of papers \cite{LopesCardoso:1998tkj,LopesCardoso:1999cv,LopesCardoso:1999fsj,LopesCardoso:2000qm}, which determined the complete off-shell background and the exact black hole Wald entropy that corrects the two derivative Bekenstein-Hawking result. In addition, the same background has been a subject of several non-renormalization theorems, stating that no possible $D$-terms can correct the entropy \cite{deWit:2010za} and that the full $\mathbb{T}$-invariant\footnote{In order to write down the $\mathbb{T}$-invariant in this case we cannot use the choice in \eqref{eq:choicephiprime} due to the vanishing of all FI parameters. Instead we use $\Phi' = \cX^0$ without loss of generality. As already discussed, different choices differ by $D$-terms, but these also vanish here.} vanishes on this background as well \cite{Butter:2014iwa},
 \be
\label{eq:ATforOSV}
	A_\mathbb{T} = 0\ .
\ee
This is rather easy to see from the explicit off-shell background, since all fields that contribute to $A_\mathbb{T}$, c.f.\ \eqref{eq:finalAWAT}, actually vanish
\be
	A_\mu = \cV_\mu{}^i{}_j = D =  R = Y^I_{i j}  = \cF^\pm_{a b} = 0\ ,
\ee
see e.g.\ \cite{Murthy:2013xpa} for a complete summary of the off-shell Bertotti-Robinson solution. The non-vanishing background fields of \cite{LopesCardoso:1998tkj,LopesCardoso:1999cv,LopesCardoso:1999fsj,LopesCardoso:2000qm} were rewritten in a particularly useful way in \cite{Sahoo:2006rp,deWit:2011gk}, which we follow here:
\be
	T^-_{01} = \i\, T^-_{23} = - w\ , \qquad v_1 = v_2 = 16\, |w|^{-2}\ ,
\ee
where $v_{1,2}$ parametrize the sizes of the AdS$_2$ and S$^2$ prefactors, respectively,
\be
	{\rm d} s^2 = v_1\, {\rm d} s^2_{AdS_2} + v_2\, {\rm d} s^2_{S^2}\ .
\ee
Going back to the discussion about gauge-fixing in section \ref{sec:setup}, in the ungauged case we now decide to use the $A$ and $D$ gauges to fix 
\be
	v_1 = v_2 = e^{- \cK_{2 \partial} (z, \bar{z})}\ , \qquad w = 4 \i\, e^{\cK_{2 \partial} (z, \bar{z})/2}\ .
\ee
As will be seen in section \ref{subsec:black holes twist}, this choice is going to be equivalent to the one we made in \eqref{eq:partialgauge} in the presence of gauging. Given the above, we arrive at
\be
	A_\mathbb{W} =  e^{\cK_{2 \partial} (z, \bar{z})}\ ,
\ee
which, together with \eqref{eq:ATforOSV}, is another justification for the {\it max}-gluing rule. One should of course remember that the precise value of $A_\mathbb{W}$ is not gauge invariant, in contrast to the on-shell action and entropy function, but importantly here it must be non-zero for the background solution to make sense.

The resulting Wald entropy for the general higher derivative case, derived in \cite{LopesCardoso:1998tkj,LopesCardoso:1999cv,LopesCardoso:1999fsj,LopesCardoso:2000qm}, was again rewritten in a more suggestive form in \cite{Ooguri:2004zv}. The full OSV formula for the on-shell action is then given by
\be
\label{eq:HDOSV}
	\cF_\text{OSV} (\phi^I, p^I) = - 8 \pi^2\, {\rm Im}\left( F (p^I + \frac{\i}{\pi}\, \phi^I; 4, 0) \right)\ ,
\ee
and the resulting entropy function is defined again as in \eqref{eq:OSVform}. Here we did not insert explicitly the Newton constant keeping the original form of the expressions, which now depend on the full prepotential $F(X^I; A_\mathbb{W}, A_\mathbb{T})$, but instead included an overall factor of $8 \pi$ that was taken out in \cite{LopesCardoso:1998tkj,LopesCardoso:1999cv,LopesCardoso:1999fsj,LopesCardoso:2000qm} as a normalization choice. 

The above formula matches the answer derived by direct evaluation of \eqref{eq:conj1} if we assume that the prepotential has no explicit powers of $\i$ (which is indeed the case for asymptotically flat compactifications). Using the {\it max}-gluing rule with $\omega = -1$, the on-shell action is given by
\be
\label{eq:HDus}
	\cF (\chi^I, p^I) = -4 \i \pi^2\,  \left( F(\kappa^{-1} (\chi^I + p^I); 4, 0 ) -  F(\kappa^{-1}( \chi^I -p^I); 4, 0) \right)\ , 
\ee
in agreement with \eqref{eq:HDOSV}, upon the identification $\phi^I = - \i \pi \chi^I$ and the reinsertion of the Newton constant in \eqref{eq:HDOSV} following the choices made in section \ref{sec:setup}. This corresponds to the {\it unrefined} limit, where the conjectured expression \eqref{eq:conj1} is correct at {\it all orders}. Due to the lack of equally exhaustive results for the higher derivative corrections of the other examples, this is the only all-order case where the general conjecture can be proven.

\subsubsection*{The topological string and the black hole partition function}
Based on the general form of on-shell action and resulting entropy function, together with the relation between the topological string and the full expansion of the (unrefined limit of the) prepotential \cite{Bershadsky:1993ta,Bershadsky:1993cx,Antoniadis:1993ze}, \cite{Ooguri:2004zv} proposed a form of the black hole partition function that was later made more preicse by \cite{Denef:2007vg}. We repeat the main formulae very briefly, in order to show that the proposed partition function is in line with part II of the present conjecture.

The unrefined topological string partition function can be written as
\be
	Z_\text{top} (t^i, g_s) = \exp \left( F_\text{top} (t^i, g_s) \right)\ ,
\ee
where $g_s$ is the string coupling constant and $t^i$ are the K\"{a}hler moduli of the underlying Calabi-Yau manifold. $F_\text{top}$ has the natural form of an expansion in $g_s$,
\be
	F_\text{top} (t^i, g_s) = \sum_h (g_s)^{2 h -2}\, F_{\text{top}, h} (t^i)\ .
\ee
Explicit calculations \cite{Bershadsky:1993ta,Bershadsky:1993cx} lead to
\be
\label{eq:ftop}
	F_\text{top} (t^i, g_s) = \frac{(2 \pi \i)^3}{6 g_s^2}\, c_{i j k} t^i t^j t^k - \frac{2 \pi \i}{24}\, c_{2, i} t^I + ...\ , 
\ee
where $c_{2, i}$ are the second Chern class numbers of the manifold, and the ellipses denote instanton corrections that are also known explicitly, see \cite{Denef:2007vg} and \cite{Murthy:2015zzy}. It is therefore natural to identify the full supergravity prepotential (at $A_\mathbb{T} = 0$) with the free energy of the topological string, leading schematically to
\be
	g_s \propto \frac{\i}{X^0}\ , \qquad t^I \propto \frac{X^i}{X^0}\ ,
\ee
see \cite{Ooguri:2004zv,Denef:2007vg} for the explicit map. As commented earlier, consistency here requires setting the magnetic charge $p^0$ to zero, such that $g_s$ remains real. The second term in \eqref{eq:ftop} is independent of $g_s$ and therefore translates to a four derivative term linear in $A_\mathbb{W}$ in the supergravity prepotential. In the present conventions, following \cite{Butter:2014iwa}, we find the corrected supergravity prepotential
\be
\label{eq:unrefinedcubicprepotential}
	F_\text{unref.} = -\frac16\, c_{i j k}\, \frac{X^i X^j X^k}{X^0} - \frac{A_\mathbb{W}}{32}\, c_{2, i}\, \frac{X^i}{X^0}\ .
\ee
It is also possible to include instanton terms of order $\exp(-g_s)$ by keeping track of how these rearrange inside the supergravity prepotential, but we are not going to pursue this subject here. We should however caution the reader that the map between the topological string free energy and the supergravity prepotential has been established under the implicit assumption that the asymptotic Minkowski space preserves supersymmetry, which is automatic here but will present a difficulty in the next subsection. 

Once the correct supergravity prepotential is identified in the unrefined limit, it only remains to evaluate the full black hole partition function, also known as the quantum entropy function (QEF). The proposal of \cite{Ooguri:2004zv,Denef:2007vg}, which was rederived in \cite{Dabholkar:2010uh,Dabholkar:2011ec} from a localization point of view, can be summarized in the following integral expression,
\be
\label{eq:finalOSVentropyfunction}
	Z_\text{OSV} (p^I, q_I) =  \int \left( \prod_{I = 0}^{n_V} {\rm d} \phi^I \right)\, e^{\cI_\text{OSV}(\phi^I, p^I, q_I)}\, Z^\text{1-loop} (\phi^I, p^I)\, Z^\text{m} (\phi^I, p^I)\ ,
\ee
where the last two pieces were initially put together into the measure factor $\mu (p^I, \phi)$ in \cite{Denef:2007vg}, but belong to two separate categories from supergravity localization point of view. Using the explicit relation between the prepotential and the topological string free energy, one can further rewrite the above expression as an integral over the K\"{a}hler moduli. It is worth noting that the 1-loop factor above was determined by \cite{Denef:2007vg} in terms of the K\"{a}hler potential and later rederived in \cite{Gupta:2015gga,Murthy:2015yfa} using the Atiyah-Singer index theorem. Due to the underlying background, the index theorem in turn simply leads to the Atiyah-Bott fixed point formula that sums over two equivalent contributions at the poles of the sphere. The measure factor in the above formula is instead giving further instanton corrections and has not been determined in full generality yet.

These gravitational results can be reproduced exactly in the dual gauge theory description via a system of D- or M-branes, \cite{Strominger:1996sh,Maldacena:1997de}. The quantity of interest is the supersymmetric index, which is naturally defined in the grand-canonical ensemble,
\be
\label{eq:susyindexosv}
	\exp \left( I_\text{CFT} (p^I, \Delta^I) \right) = {\rm Tr} (-1)^F\, e^{-\beta H}\ , 
\ee 
The holographic identification goes through the transformation of the above index to the grand-canonical ensemble, or alternatively the inverse transform of the microcanonical expression in \eqref{eq:finalOSVentropyfunction}, see again \cite{Ooguri:2004zv,Denef:2007vg} and references thereof. Due to supersymmetry, the index is a protected quantity and can be computed at any point on the moduli space of the dual field theory, which greatly facilitates the successful holographic match. 

The above results are in full agreement with the general discussion in part II of the present conjecture. Spelling out \eqref{eq:conj2} in the present case leads to
\be
	Z (p^I, q_I) := \int \left( \prod_{I = 0}^{n_V} {\rm d} \chi^I \right)\,  e^{\cI (\chi^I, p^I, q_I)}\, Z^\text{UV} (\chi^I, p^I)\ ,
\ee
with 
\be
	\cI (\chi^I, p^I, q_I) = - \cF (\chi^I, p^I) - \frac{8 \i \pi^2}{\kappa^2}\, \chi^I q_I\ ,
\ee
and $\cF (\chi^I, p^I)$ determined in \eqref{eq:HDus}. We find a precise agreement with \eqref{eq:finalOSVentropyfunction} upon the identification $\phi^I = - \i \pi \chi^I$ and the relation
\be
	Z^\text{UV} (\chi^I, p^I) = Z^\text{1-loop} (- \i \pi \chi^I, p^I)\, Z^\text{m} (- \i \pi \chi^I, p^I)\ ,
\ee
which can be considered as a particular constraint on the general form of the UV completion of the gravitational Nekrasov partition function in the unrefined limit $\omega = -1$. We can therefore conclude that both part I and part II of the general conjecture are in agreement with the known results for asymptotically flat BPS black holes, which sit precisely at the unrefined limit of the main formulae \eqref{eq:conj1} and \eqref{eq:conj2}.

\subsection{Static/under-rotating non-BPS black holes}
\label{subsec:under}
We now consider the somewhat peculiar example of the so called under-rotating \cite{Bossard:2012xsa}, or slow rotating \cite{Chow:2014cca}, non-BPS black holes in flat space. These solutions have been considered and analysed in multiple references, including \cite{LopesCardoso:2007qid,Gimon:2007mh,Bena:2009ev,DallAgata:2010srl,Bossard:2012xsa} and references thereof. The most general class of these solutions was described in \cite{Bossard:2012xsa} and identified in \cite{Chow:2014cca} within the general black hole solutions of the STU model. These black holes are extremal and in general rotating, with the near-horizon geometry being a non-trivial fibration of AdS$_2$ and S$^2$, but admit a smooth limit to a static extremal non-BPS solution with a near-horizon AdS$_2 \times$S$^2$. Within ungauged supergravity one can see that supersymmetry is only restored asymptotically for the underlying Minkowski vacuum, and broken otherwise. Instead, \cite{Hristov:2012nu} showed that a special type of gauged supergravity that only switches on the FI parameter $g_0$ (with the same cubic prepotential \eqref{eq:usualcubicprepotential}) can accommodate for the same solutions due to the vanishing scalar potential, but results in a change of the supersymmetry properties of this background. In this {\it flat gauged} supergravity it is the near-horizon geometry that preserves supersymmetry via a topological twist (which requires that $g_0$ is tuned precisely to the value $-1/p^0$), while the full black hole flow and the asymptotic Minkowski spacetime manifestly break supersymmetry \cite{Hristov:2012nu,Gnecchi:2013mja}. More recently, the general rotating horizons were shown to belong to the class of twisted rotating attractors in gauged supergravity, \cite{Hristov:2018spe}, and their entropy function is therefore described by the {\it A}-gluing rule as shown explicitly in appendix B of \cite{Hosseini:2019iad}. 

Just like ungauged supergravity, the flat gauged supergravity can also be obtained by higher dimensional compactifications, e.g.\ from five dimensions, via a Scherk-Schwarz reduction \cite{Hristov:2012nu,Hristov:2014eba} and therefore we consider it as an alternative description of the same class of under-rotating solutions. One can think of this as a simple trick that allows us an easier description of the higher derivative near-horizon background. In fact the other possibility, using Sen's entropy function formalism to analyze the higher derivative corrections in ungauged supergravity, was accomplished in \cite{Sahoo:2006rp} in the static limit. The resulting conclusion that the $\mathbb{W}$-invariant is not enough to capture all corrections, in contrast to the BPS solutions in the previous subsection, was later understood by \cite{Banerjee:2016qvj} after the discovery of the $\mathbb{T}$-invariant \cite{Butter:2013lta} and its identification from five-dimensional compactifications \cite{Butter:2014iwa}. It is this particular example that prompted the authors of \cite{Saraikin:2007jc} to suggest a relation with the refined form of the Nekrasov partition function, ultimately inspiring the present conjecture.

\subsubsection*{Two derivative solutions}
Once we specify to the flat gauged supergravity discussed above, 
\be
	g_0 \neq 0\ , \qquad \qquad g_i = 0\ , \qquad \forall i \neq 0\ ,
\ee
the general rotating near-horizon solutions of interest can be written in the formalism of \cite{Hristov:2018spe}. In more detail, the solutions in \cite{Hristov:2018spe} are actually based on the assumption of a symmetric scalar manifold for the vector multiplet scalars $z^I$, which is satisfied by the cubic prepotentials considered here \eqref{eq:usualcubicprepotential} under an additional condition for the intersection numbers $c_{i j k}$. In these cases one can define the so called quartic invariant $I_4$ and make use of the symplectic formalism to rewrite the differential BPS equations into algebraic equations involving the quartic invariant and the basic symplectic vectors of the solutions, e.g.\ the vector of electromagnetic charges $\Gamma = \{p^I; q_I \}$. Without going into further details, which can be found in e.g.\ \cite{Bossard:2012xsa,Hristov:2018spe,Hosseini:2019iad}, it is intuitively useful to compare the form of the black hole entropy for the BPS black holes considered in the previous subsection,
\be
	S = \frac{\pi}{G_N^{(4)}}\, \sqrt{I_4 (\Gamma)}\ ,
\ee
with $I_4 (\Gamma)$ a particular quartic function of the electro-magnetic charges, and the form of the entropy for the under-rotating branch we consider here,
\be
\label{eq:underroti4entropy}
	S = \frac{\pi}{G_N^{(4)}}\, \sqrt{-I_4 (\Gamma) - \cJ^2}\ ,
\ee
where $\cJ$ is the angular momentum, bounded from above by the value $-I_4 (\Gamma)$ (leading to the name {\it under-rotating}). One can see that even in the static limit the solutions are different due to the need of different overall sign of $I_4 (\Gamma)$, which in turn requires at least one of the charges to be of opposite sign.

It was shown in \cite{Hosseini:2019iad} that the corresponding entropy function can be derived the {\it A}-gluing rule, which follows from the fact that supersymmetry is preserved via the twisting condition $g_0 p^0 = -1$.\footnote{It is therefore crucial that the magnetic charge $p^0$ is non-vanishing. This might seem odd for asymptotically flat solutions, and indeed this condition can be lifted upon restoring full electromagnetic duality with the help of the embedding tensor formalism, see e.g.\ \cite{deWit:2011gk}. In order not to introduce magnetic gaugings here we just assume $p^0 \neq 0$.} A more detailed explanation and justification of the {\it A}-gluing can be found in section \ref{subsec:black holes twist}, which is the better known case of a black hole with a twist. Here we just repeat the main conclusion that the two derivative entropy function is given by
\be
\label{eq:entropyfununderrot}
	\cI (\chi^I, \omega, p^I, q_I, \cJ) = \frac{\i \pi}{2 \omega G_N^{(4)}} \left( F(\chi^I+\omega p^I) - F(\chi^I - \omega p^I) \right) - \frac{\i \pi}{G_N^{(4)}} (\chi^I q_I - \omega \cJ)\ ,
\ee
under the constraints $\chi^0 = 1/g_0 = -p^0$, where we need to use the explicit cubic prepotential \eqref{eq:usualcubicprepotential}.\footnote{Note that here we use the opposite overall sign for the two derivative prepotential with respect to \cite{Hosseini:2019iad}, which results in flipping some signs in the so called quartic invariant $I_4$ used for constructing the explicit solutions according to \cite{Hristov:2018spe}.} The black hole entropy \eqref{eq:underroti4entropy} is precisely reproduced upon extremizing the above function, in agreement with \eqref{eq:entropy}. The critical values of the gluing parameters $X^I_{(1), (2)}$ are proportional to the near-horizon background values of the on-shell sections at the poles of the sphere,
\be
	X^I_\text{SP} (z) = \i\, X^I_{(1)} \Big|_\text{crit.}\ , \qquad  X^I_\text{NP} (z) = \i\, X^I_{(2)} \Big|_\text{crit.}\ ,
\ee
in agreement with the analogous answer for the twisted black holes in AdS$_4$, c.f.\ section \ref{subsec:black holes twist}.

\subsubsection*{Higher derivatives, the static limit and the refined topological string}
Looking at the higher derivative case, we again opt to leave for section \ref{subsec:black holes twist} the general prediction for the entropy function and the explicit form of the static off-shell twisted near-horizon background in gauged supergravity following \cite{Hristov:2016vbm}. Instead we are going to focus directly on the available explicit results concerning the embedding of the non-BPS black holes in question in string theory, summarizing the work of \cite{Butter:2014iwa} and \cite{Sahoo:2006rp,Banerjee:2016qvj}.

Shortly after the discovery of the $\mathbb{T}$-invariant in \cite{Butter:2013lta}, \cite{Butter:2014iwa} showed that the straightforward circular compactification of five-dimensional higher derivative supergravity with the Weyl$^2$ invariant \cite{Hanaki:2006pj} produces a special combination of the $\mathbb{W}$ and $\mathbb{T}$ invariants in four dimensions. This allowed the authors to determine the four derivative prepotential that generalizes the topological string prediction \eqref{eq:unrefinedcubicprepotential} from the previous subsection,
\be
\label{eq:refinedcubicprepotential}
	F_\text{ref.} = -\frac16\, c_{i j k}\, \frac{X^i X^j X^k}{X^0} - \frac{(3 A_\mathbb{W}+A_\mathbb{T})}{96}\, c_{2, i}\, \frac{X^i}{X^0}\ ,
\ee
where, as in the previous subsection, the coefficients $c_{ijk}$ are the Calabi-Yau intersection numbers, while $c_{2, i}$ are the second Chern class numbers, \cite{Bershadsky:1993ta,Bershadsky:1993cx}.

Based on the above prepotential and Sen's entropy function \cite{Sen:2005wa} applied to the resulting higher derivative Lagrangian, \cite{Banerjee:2016qvj} resolved an earlier puzzle that was pointed out in \cite{Sahoo:2006rp} before the discovery of the $\mathbb{T}$-invariant. Namely, \cite{Sahoo:2006rp} discovered, based on the five dimensional uplift of the BPS and non-BPS solutions discussed here, that the static BPS and non-BPS black holes receive different higher derivative corrections even though their two derivative entropies are precisely equal upto flipping the sign of $I_4 (\Gamma)$. At the time this lead to a puzzle when considering only the $\mathbb{W}$-invariant in four dimensions, which could not account for the different corrections in these two cases. After the discovery of the $\mathbb{T}$-invariant and the explicit reduction leading to \eqref{eq:refinedcubicprepotential}, \cite{Banerjee:2016qvj} showed that the BPS solutions discussed in the previous section only receive corrections from the $\mathbb{W}$-invariant, while the non-BPS branch receives corrections from both the $\mathbb{W}$ and $\mathbb{T}$-invariants. 

We can now understand the resolution of the original puzzle in our formulation. When looking at the near-horizon geometry in flat gauged supergravity with $g_0 \neq 0$, we need to use the {\it A}-gluing. Referring for more details to section \ref{subsec:black holes twist} based on the results in \cite{Hristov:2016vbm},\footnote{We need to correct a misleading interpretation presented in section 5.1 of \cite{Hristov:2016vbm}, claiming that the flat gauged and ungauged supergravity descriptions become inequivalent in the presence of higher derivatives, on the basis of the fact that the radii of the AdS$_2$ and S$^2$ factors in the near-horizon geometry are no longer equal. In fact the exact same phenomenon is observed in \cite{Sahoo:2006rp,Banerjee:2016qvj} on the basis of Sen's entropy function with higher derivatives in the ungauged model. We therefore remain with the freedom of considering the under-rotating branch in the presence of higher derivatives in both the ungauged and flat gauged supergravity.} here we reproduce the resulting higher derivative on-shell action in the limit of vanishing angular momentum $\cJ = 0$, which corresponds to taking $\omega \rightarrow 0$ in this case,
\be
\label{eq:statictwistedsphericalHD}
	\cF (p^I, \chi^I) = -8 \i \pi^2\, \left(\kappa^{-1}\, p^I F_I(\kappa^{-1} \chi^I; 1, 1) + 2  F_{A_\mathbb{W}} (\kappa^{-1} \chi^I; 1, 1) -2  F_{A_\mathbb{T}} (\kappa^{-1} \chi^I; 1, 1) \right)\ ,
\ee
with $\chi^0 = 1/g_0 = -p^0$. Since the prepotential \eqref{eq:refinedcubicprepotential} leads to
\be
	F_{A_\mathbb{W}} (\kappa^{-1} \chi^I; 1, 1) = 3 F_{A_\mathbb{T}} (\kappa^{-1} \chi^I; 1, 1)  = \frac{1}{32}\, c_{2, i}\, \frac{\chi^i}{p^0}\ ,
\ee
it is clear that both invariants give non-zero contributions, combining to give
\be
	\cF (p^I, \chi^I) = -\frac{4 \i \pi^2}{3 p^0\, \kappa^2} \left( c_{i j k} (\chi^i + 3 p^i) \chi^j \chi^k + \frac14 \kappa^2\, c_{2, i} (2 \chi^i + p^i)  \right)\ .
\ee
The resulting entropy function after Legendre transforming this on-shell action is indeed consistent with the explicit black hole entropy presented in \cite{Sahoo:2006rp,Banerjee:2016qvj}.\footnote{Note that the duality frame considered in \cite{Sahoo:2006rp,Banerjee:2016qvj} corresponds to a vanishing $p^0$ charge. Therefore the comparison with the natural flat gauged supergravity solution passes through a symplectic rotation of either of the solutions to the opposite duality frame that preserves the same prepotential.}

It is now tempting to speculate that the prepotential \eqref{eq:refinedcubicprepotential} can be related to the free energy of the {\it refined} topological string, which naturally generalizes the expression in \eqref{eq:ftop} to allow for one additional equivariant parameter on top of $g_s$. In this case the corresponding refined free energy would have an expansion much like the one in \eqref{eq:1} with both $F^{(1,0)}$ and $F^{(0,1)}$ proportional to $c_{2, i} t^i$, generalizing \eqref{eq:ftop} in a natural way. However, we should caution against this naive approach due to the non-BPS nature of the black holes we considered. Whether one considers them in ungauged or in flat gauged supergravity, these solutions are only preserving supersymmetry either asymptotically or strictly at the horizon, but nowhere along the flow. Therefore we can no longer argue that the corresponding partition function is protected against $D$-terms. This is equivalent with the statement that the refined topological string free energy on compact Calabi-Yau manifolds can depend not only on the K\"{a}hler, but also on the complex structure moduli, see e.g.\ \cite{Aganagic:2011mi,Aganagic:2012si,Alexandrov:2019rth} and references therein. One could however still argue that there exists a particular point on the K\"{a}hler moduli space where the dependence on the complex moduli actually disappears, as suggested by the flat gauged supergravity picture where the near-horizon is indeed protected from $D$-terms.

On a related note, we can now construct the grand-canonical $Z(p^I, \chi^I, \omega)$ and microcanonical $Z(p^I, q_I, \cJ)$ partition functions using the on-shell action \eqref{eq:statictwistedsphericalHD} and the prepotential \eqref{eq:refinedcubicprepotential} in a straightforward way as outlined in part II of the conjecture (see again section  \ref{subsec:black holes twist} for the general formulae that follow from the {\it A}-gluing). In the holographically dual theory, based on the same $D$-brane or $M$-brane systems relevant for the BPS black holes in the previous subsection, one should instead consider the refined index (with angular momentum), also called {\it spin character} index \cite{Aganagic:2012si,Alexandrov:2019rth}, that generalizes \eqref{eq:susyindexosv} with the generator of rotations $L_3$,
\be
	\exp \left( I^\text{ref.}_\text{CFT} (p^I, \Delta^I) \right) = {\rm Tr} (-1)^F\, e^{-\beta H} e^{\i\, \varepsilon L_3}\ . 
\ee
Again, the main problem of this quantity is that it is not protected and in general does depend on the position on the moduli space of the theory. We can similarly expect that there exists a special point where one can trust the duality, but it is clear that this case begs for a more careful analysis and will be explored elsewhere.

\section{Asymptotically AdS$_4$ solutions}
\label{sec:AdS}
Here we focus on a number of interesting asymptotically locally AdS$_4$ solutions. The two-derivative prepotentials (with a known higher-dimensional origin) that allow for a supersymmetric AdS$_4$ vacuum are usually of the square-root type as suggested in table \ref{tab:2}, and we are going to stick to this case (and its truncations to smaller sectors) in the explicit examples we discuss below.

\subsection{Universal solutions in the minimal theory}
\label{subsec:universal}
As already outlined in the introduction, the case of minimal supergravity with $n_V = 0$ leads to a considerable technical simplification, while in the same time offering important holographic insights. A major simplification in this limit occurs already without imposing any restrictions on the background geometry - the $D$-terms of \cite{deWit:2010za} automatically vanish in the off-shell formalism. Moreover, due to homogeneity of the $F$-terms, the general form of the prepotential \eqref{eq:1} in the limit $n_V = 0$ becomes
\be
\label{eq:prepotmin}
	F (X^0; A_\mathbb{W}, A_\mathbb{T}) = -\i\, \sum_{m,n = 0}^\infty f_{(m,n)} (X^0)^{2 (1-m-n)}\, (A_\mathbb{W})^m\, (A_\mathbb{T})^n\ ,
\ee
where $f_{(m,n)}$ are a set of arbitrary constants, e.g.\ we can set $f_{(0,0)} = 1, f_{(1,0)} = c_1, f_{(0,1)} = c_2$ and so on. Without loss of generality we can also set $g_0 = 2$, which in turn sets the AdS$_4$ length scale to $L = 1$. The four-derivative case, where only the first three coefficients are non-zero,\footnote{Alternatively, we can use the definition \eqref{eq:compositeA} and write the four-derivative prepotential in minimal supergravity as $ F(X^0; \cA) = -\i\, (X^0)^2 - \i\, \cA$.}
\be
\label{eq:RHCBprepot}
	F (X^0; A_\mathbb{W}, A_\mathbb{T}) = - \i (X^0)^2 -\i c_1\,  A_\mathbb{W} - \i c_2\, A_\mathbb{T}\ ,
\ee
was considered in \cite{Bobev:2020egg,Bobev:2020zov,Bobev:2021oku} and the general on-shell action for an arbitrary supersymmetric solution with fixed points under the canonical isometry was conjectured in \cite{Bobev:2021oku} and proven in \cite{Genolini:2021urf} (by generalizing a similar proof in the two-derivative case \cite{BenettiGenolini:2019jdz}). Written in the present conventions,\footnote{\label{ft:25}Note that \cite{Bobev:2020egg,Bobev:2020zov,Bobev:2021oku} use Euclidean supergravity and different normalization of the higher derivative invariants, resulting in a number of sign and prefactor changes. For the sake of direct comparison we can adjust the final answers by the map $2\, c_1^\text{there} = - c_1^\text{here}$ and $2\, c_2^\text{there} =  c_2^\text{here}$.} the four-derivative answer is given by
\be
\label{eq:fourdermin}
	\cF = \sum_{\sigma \in M_4} \frac{\pi^2}{\omega_{(\sigma)}}\, \left( (\kappa^{-2} + 4\, c_2)\, (1+\omega_{(\sigma)})^2 + 4\, c_1\, (1-\omega_{(\sigma)})^2  \right)\ ,
\ee
which is exact agreement with the result of the general conjecture \eqref{eq:conj1},
\be
\label{eq:mingluing}
	\cF = \sum_{\sigma \in M_4} \frac{4 \i \pi^2\, F(\kappa^{-1} X^0_{(\sigma)}; (1-\omega_{(\sigma)})^2, (1+\omega_{(\sigma)})^2)}{\omega_{(\sigma)}}\ , \quad X^0_{(\sigma)} = \frac{(1 + \omega_{(\sigma)})}{2}\ ,
\ee
applied to the above prepotential. As discussed in the introduction, it can be obtained both from the $\it identity$ and from the {\it S}-gluing, since $\cF(\text{KN-AdS}) = 2\, \cF(S^3_\text{sq.})$ in this minimal limit. It is in fact clear that the gluing trivializes by plugging in $X^0_{(\sigma)} = (1 + \omega_{(\sigma)})/2$ directly inside \eqref{eq:mingluing} and one can introduce the above formula as in \cite{BenettiGenolini:2019jdz,Genolini:2021urf} without any need to specify a gluing rule. 

The general conjecture \eqref{eq:conj1}, specialized to the $n_V = 0$ case, therefore proposes that the on-shell action for the holographic sphere and Kerr-Newman-like black hole backgrounds takes the form of \eqref{eq:mingluing} for any number of derivatives, generalizing \eqref{eq:fourdermin} to allow for the full expansion in \eqref{eq:prepotmin}. For the case of pure AdS$_4$ with round S$^3$ boundary (such that $\omega (S^3) = 1$) the infinite derivative expansion was already evaluated in section 7.1 of \cite{Bobev:2021oku} and we find that it agrees with the form proposed above.\footnote{One should be careful not to look up directly the final on-shell result of \cite{Bobev:2021oku}. Due to a different choice for the insertion of the Newton constant via the gauge-fixing procedure, the results only appear similar but not precisely equal. They become equivalent if we insist on the choices made in section \ref{sec:setup} here.} The careful discussion of this background is postponed to the next subsection. Note that the results in \cite{Bobev:2020egg,Bobev:2020zov,Bobev:2021oku,Genolini:2021urf} cover also the backgrounds with fixed two-submanifolds, which we instead discuss as a particular limit in section \ref{subsec:black holes twist}.

It is instructive to follow more closely the faith of the off-shell section $X^0$ and its on-shell avatar $X^0 (z)$ in this very simple case.\footnote{Clearly in this case there are no physical scalars $z^I$, we only keep the notation $X^0 (z)$ to distinguish the off-shell and the on-shell quantities.} Using the fact that $g_0 = 2$, \eqref{eq:partialgauge} and the definition \eqref{eq:K-pot} lead immediately to
\be
	X^0 (z) = \frac12\ , \qquad e^{-\cK_{2 \partial} (z, \bar{z})} = 1\ ,
\ee
while applying the relation \eqref{eq:offtoon} to the off-shell sections leads to
\be
	X^0 = \frac12\ .
\ee
This is in exact agreement with the answer presented in \cite{Bobev:2021oku} after taking in account the rescaling \eqref{eq:rescaling}, and is the background value of $X^0$ for any possible solution. This leads us to the following important remark: it does not in general make sense to directly relate the background value of the sections $X^I$ to the respective gluing rules that give rise to the evaluation of the on-shell action as proposed by \eqref{eq:conj1}. In the example at hand one should rather see the  $X^0_{(\sigma)}$ inside the building block as a dummy variable. In the other examples with $n_V \neq 0$ we are going to see some specific relations between the values of the sections $X^I (z)$ at the fixed points and the respective gluing rules, but there does not appear to exist a general rule.\footnote{This was already observed and remarked upon in \cite{Hosseini:2019iad} for the two-derivative case.} The situation with the composite fields $A_\mathbb{W}, A_\mathbb{T}$ is better since their explicit values on the fixed points are proportional to the respective gluing rule, but one should always be mindful that the actual background values are gauge-dependent, while physical observables such as the on-shell action and entropy function are gauge invariant.

Finally, we should note that the minimal supergravity case is also very special from the point of view of part II of the conjecture since there only remains a single integration over the Coulomb branch parameters (counting $\omega$ among them) in \eqref{eq:conj2}. It would therefore be interesting to explore the resulting {\it universal} partition functions that will likely exhibit important general features potentially useful in understanding the general matter-coupled objects.

\subsection{The holographic three-sphere}
\label{subsec:sphere}
Here we are going to discuss the backgrounds corresponding holographically to the S$^3$ partition function \cite{Kapustin:2009kz,Hama:2011ea}. This class of backgrounds were not discussed explicitly in \cite{Hosseini:2019iad} and therefore we need to start summarizing the basic facts about the two-derivative solution before discussing the higher-derivative generalization.  Note that the known backgrounds corresponding to squashed spheres were already covered in the previous subsection,\footnote{It is straightforward to see that the corresponding gluing rule in table \ref{tab:1}, when applied to $n_V = 0$, leads to \eqref{eq:mingluing} with a single fixed point, where the deformation parameter is identified with the squashing parameter $\omega = \sqrt{b}$ for the $\U(1)\times\U(1)$ squashing \cite{Martelli:2011fu}, and $\omega = 2 s^2 - 1 + 2 s\, \sqrt{s^2-1}$ for the $\SU(2)\times\U(1)$ squashing \cite{Martelli:2011fw}. The resulting answer is in agreement with the on-shell actions derived in these references.} and it is still an open problem to construct these backgrounds in non-minimal supergravity already at two derivatives. As postulated in part I of the conjecture such solutions must exist, but in their absense we are going to mostly focus on the round S$^3$ case that exhibits the NS limit of the general building block \eqref{eq:conj1}. The round S$^3$ is a conformally flat manifold, therefore the special solution we consider is just the maximally supersymmetric AdS$_4$ vacuum with S$^3$ slicing. We also give more details on its corresponding Coulomb branch solutions, discovered in \cite{Freedman:2013oja}.

\subsubsection*{Two derivative solutions}
Let us here consider the standard two-derivative example of the electrically gauged STU model following from the Cartan truncation of maximal 4d $\cN=8$ supergravity and in turn from the compactification of 11d supergravity on S$^7$,
\be
\label{eq:2dSTUprepot}
	F_{2 \partial} = -2 \i\, \sqrt{X^0 X^1 X^2 X^3}\ , \quad g_0 = g_1 = g_2 = g_3 = 1\ , 
\ee
which results in the AdS$_4$ length scale $L = 1/\sqrt{2}$. We are going to discuss the off-shell fields on the AdS$_4$ background below, and here instead focus on the simpler task of describing the on-shell background. AdS$_4$ is in fact the unique maximally supersymmetric background \cite{Hristov:2009uj} of the theory defined by the above prepotential and gauging, and is characterized by a constant negative curvature, vanishing background gauge fields and physical scalars fixed at the extremum of the resulting scalar potential. Imposing the gauge-fixing choice \eqref{eq:partialgauge}, the AdS$_4$ vacuum fixes
\be
	X^I (z) = \frac14\ , \forall I\ , \quad e^{-\cK_{2 \partial} (z, \bar{z})} = \frac{1}{2}\ , \quad X^I = \frac{\sqrt{2}}4\ , \forall I\ .
\ee
 In the case of Euclidean AdS$_4$ with a round S$^3$ slicing, Freedman and Pufu \cite{Freedman:2013oja} found a more general half-BPS background corresponding to deformations of the dual theory encoding the most general choice of $\U(1)_R$ symmetry, preserving the (Euclidean) supersymmetry algebra $\OSp(2|2) \times \SU(2)$. The supergravity solutions are specified by running scalar fields and a radial flow with a gradually shrinking S$^3$ size, while keeping the background gauge fields vanishing. In the language introduced above, these are precisely the Coulomb branch solutions corresponding to the holographic round S$^3$ special solution. 

Using the explicit knowledge of the Killing spinors and their bilinears (for the special solutions at the origin of the moduli space), \cite{BenettiGenolini:2019jdz} showed that the canonical isometry near the fixed point takes the form
\be
	\xi = \partial_{\varphi_1} + \partial_{\varphi_2}\ ,
\ee
for the case of round holographic three-sphere, and more generally
\be
	\xi = \partial_{\varphi_1} + \omega \partial_{\varphi_2}\ ,
\ee
for the squashed case with $\omega > 0$, based on the solution of \cite{Martelli:2011fu},\footnote{We remind the reader that we decided to follow the convention of automatically fixing $s=+$ of the first (here unique) fixed point. In the cited references one can find both chiralities, which in turn also means both signs for $\omega$.} in accordance with the suggested gluing rule. If we evaluate the on-shell action from \eqref{eq:conj1} using the {\it S}-gluing in table \ref{tab:1} applied to the explicit prepotential and FI gaugings above, we find the following answer:
\be
	\cF (S^3_\text{sq}) = \frac{\pi\, (1+\omega)^2\, \sqrt{\chi^0 \chi^1 \chi^2 \chi^3}}{\omega\, G_N^{(4)}}\ , \qquad \sum_{I = 0}^3 \chi^I = 1\ . 
\ee 
When discussing the round case, based on the above results we should further restrict to the vanishing squashing limit,
\be
	\omega ( S^3) = 1\ .
\ee
The answer is precisely matching the one derived in \cite{Freedman:2013oja} by an explicit calculation taking in account the principles of holographic renormalization \cite{Klebanov:1999tb,Skenderis:2002wp,Bianchi:2001kw}. As pointed out above, the general answer for arbitrary $\omega > 0$ is in agreement with the known results in minimal supergravity, but the corresponding Coulomb branch solutions for a general squashing have not been constructed. However, it is worth pointing out that the holographically dual large $N$ calculation was undertaken in \cite{Hosseini:2019and} and the form of the on-shell action presented above is in exact agreement. This gives us confidence that the general solutions exhibitng this on-shell action truly exist.

Let us now comment more on the so-called $\cF$-extremization, which is the dual field theoretical statement that the exact superconformal symmetry extremizes the sphere partition function \cite{Jafferis:2010un,Jafferis:2011zi}. Applying the same logic to the two-derivative answer above, we can easily find the extremum of $\cF (S^3_\text{sq})$ under the constraint $ \sum_{I = 0}^3 \chi^I = 1$:
\be
	\chi^I \Big|_\text{crit.} = \frac14\ , \, \forall I\ , \qquad \omega \Big|_\text{crit.} = 1\ . 
\ee
As expected, we indeed recover precisely the special solution corresponding to pure AdS$_4$ and therefore restore superconformal symmetry in exact agreement with the field theory interpretation. Note here that one is free to consider the squashed cases with arbitrary $\omega$ while still recovering the same extremum for the Coulomb branch parameters. This recovers the minimal supergravity solutions with squashing \cite{Martelli:2011fu,Martelli:2011fw}, but does not restore full conformal symmetry. 

We should note that the explicit on-shell action above, as well as the full higher derivative expression in \eqref{eq:conj1}, is invariant under the inversion $\omega \rightarrow 1/\omega$ when using the {\it S}-gluing rule. This must be the case by the simple consideration that the holographic S$^3$ background possesses only a single fixed point, and the original definition of $\omega$ in \eqref{eq:fixedpointvector} is arbitrary in the choice of parameters $\varepsilon_{1,2}$. Instead this symmetry is broken for the black hole spacetimes due to the appearance of magnetic charges, ultimately allowing one to also explore the limit $\omega = 0$ in some cases. It is therefore immediate to see that the resulting on-shell action, with arbitrary number of derivatives, remains extremized precisely at $\omega = 1$ which gives back the round three-sphere. This is rather natural in supergravity as it corresponds to pure AdS$_4$, i.e.\ the proposed gluing leads to a set of self-consistent results. This gives us confidence to claim that we have presented the correct {\it S}-gluing rule.

\subsubsection*{Higher derivatives and the NS limit}
Here we consider only the round sphere case, $\omega = 1$. According to the general higher-derivative expression for the gravitational block \eqref{eq:conj1}, in this limit the whole $\mathbb{W}$ tower of terms is evaluated to zero. It is easy to see how this comes out from the off-shell formalism. Just by imposing maximal symmetry on the AdS$_4$ background leads to the vanishing of all fields with spacetime indices,
\be
	A_\mu = \cV_\mu{}^i{}_j = W_\mu^I = T^\pm_{a b} =  \cG^\pm_{a b} = 0\ , 
\ee
which from the definition of $A_\mathbb{W}$ immediately leads to
\be
	A_\mathbb{W}= 0\ .
\ee
In addition it is clear that the Weyl tensor vanishes on AdS$_4$, which is another way of understanding why the full $\mathbb{W}$-invariant actually evaluates to zero on this background. This is clearly valid at the superconformal point, i.e.\ for the special holographic round S$^3$ background, but we expect the Freedman-Pufu solutions to not excite any vectors and tensors such that the above equations hold again. This expectation also agrees with the conjectural approach in section 7.2 of \cite{Bobev:2021oku} where a partially off-shell evaluation of the pure AdS$_4$ on-shell action was proposed to agree with the result from the evaluation of the higher-derivative corrections on the Freedman-Pufu solutions. The fact that half of the possible $F$-terms vanish for the holographic round S$^3$ partition function was independently argued on very general grounds in \cite{Binder:2021euo} based on the background symmetry superalgebra, where the authors showed that all $D$-terms automatically vanish as well. 

The evaluation of the $\mathbb{T}$-invariant requires a few more technical steps as it also involves the values of the physical and auxiliary scalars, c.f.\ \eqref{eq:finalAWAT}. Here we can use the power of full supersymmetry (as we focus on pure AdS$_4$), where all the needed information is determined by the simple requirement that all different terms in the $Q$-supersymmetry variations vanish identically. This fixes the following background values:\footnote{Strictly speaking, these results seem to appear here for the first time. They are however straightforward to obtain by combining the minimal supegravity results in App. E of \cite{Bobev:2021oku} with the matter-coupled Lorentzian supersymmetric conditions in \cite{Hristov:2016vbm}.}
\be
	D = 0\ , \quad R = 12 |g_I X^I|^2 = 12 |X|^2\ , \quad Y_{i j} = g_I Y^I_{i j} = 2 \i\,  |g_I X^I|^2\, \varepsilon_{i k} \sigma^3{}^k{}_j \, .
\ee
Imposing the conditions \eqref{eq:offtoon} and \eqref{eq:partialgauge} leads to the complete simplification of the composite field $A_\mathbb{T}$,
\be
	A_\mathbb{T} =  e^{\cK_{2 \partial} (z, \bar{z})}\ ,
\ee
which holds for an arbitrary choice of prepotential and gauging that allow for an AdS$_4$ vacuum. We should note that we have not yet evaluated the individual sections $X^I (z)$ at the higher-derivative level, and it is in general impossible to do so without explicitly specifying the prepotential. The values of the composite fields $A_{\mathbb{W}, \mathbb{T}}$ are exactly reversed with respect to the asymptotically flat black holes in section \ref{subsec:osv2}, as expected from the fact that here $\omega = 1$ instead of $-1$.

These results are in complete agreement with the proposed Nekrasov-like building block \eqref{eq:conj1}, for which the value $\omega = 1$ is very special and corresponds to the Nekrasov-Shatashvili limit of the full formal expansion \eqref{eq:1}. The {\it S}-gluing rule at $\omega =1$ gives the general prediction for the on-shell action
\be
\label{eq:roundS3HDaction}
	\cF(S^3, \chi^I, \omega=1) =4 \i \pi^2\, F(2 \kappa^{-1}\, \chi^I; 0, 4)\ ,
\ee
under the constraint $g_I \chi^I = 1$, where the only surviving higher derivative contributions from the general form of the prepotential in \eqref{eq:1} are from the coefficients $F^{(0,n)}$. Notice that the four derivative coefficient, $F^{(0,1)}$, is of degree zero in $\chi^I$, and therefore does not change the critical point under $\cF$-extremization with respect to the two derivative answer. Every other coefficient starting from six derivatives has a non-zero degree and therefore can potentially shift the extremum, but the shift is of order at least $(G_N^{(4)})^2$. 

Notice that the proposed form of the full higher-derivative on-shell action for the holographic S$^3$ background is also in exact agreement with the explicit results for the infinite-derivative expansion in minimal supergravity and the matter-coupled four-derivative prepotential considered in sections 7.1 and 7.2 of \cite{Bobev:2021oku}, respectively. These results should be considered conjectural due to the lack of proper treatment of the higher-derivative holographic renormalization procedure, but neverthelsss give further evidence for the credibility of the proposed on-shell action. Based on the holographic match with the ABJM theory and the explicit results of \cite{Nosaka:2015iiw}, it was further proposed in \cite{Bobev:2021oku} that the explicit four-derivative prepotential for the STU model takes the form
\be
	F^{(0,1)} \propto -\i\, \frac{\sum_{I=0}^3 (X^I)^2}{\sqrt{X^0 X^1 X^2 X^3}}\ ,
\ee
and it would be interesting to holographically test this proposal further against some of the answers from the black hole backgrounds discussed here.

\subsubsection*{Squashing and the Cardy limit}
For completeness, let us also reproduce here the prediction for the complete expression of the on-shell action of the general holographic squashed sphere background, following from \eqref{eq:conj1} and the {\it S}-gluing rule,
\be
	\cF(S^3, \chi^I, \omega) = \frac{4 \i \pi^2}{\omega}\, F (\kappa^{-1}\, (1+\omega) \chi^I; (1-\omega)^2, (1+\omega)^2)\ ,
\ee
under the constraint $g_I \chi^I = 1$, leading back to \eqref{eq:roundS3HDaction} in the limit $\omega = 1$ for the prepotential \eqref{eq:1}. In the limit of minimal supergravity this expression 
reproduces \eqref{eq:mingluing} at a single fixed point, as expected, and is therefore confirmed at four derivative level by the results in \cite{Bobev:2020egg,Bobev:2020zov,Bobev:2021oku}.

We can now consider another interesting limit, $\omega \rightarrow 0$, known as the Cardy limit. It geometrically corresponds to infinite squashing, due to the symmetry $\omega \rightarrow 1/\omega$ ,meaning that it is the opposite of the round S$^3$ discussed above. In this limit we find
\be
\label{eq:cardysquashedsphere}
	\cF_\text{Cardy} (S^3, \chi^I, \omega) = \lim_{\omega \rightarrow 0} \cF(S^3, \chi^I, \omega) = \frac{4 \i \pi^2}{\omega}\, F (\kappa^{-1}\, \chi^I; 1, 1)\ ,
\ee 
with $g_I \chi^I = 1$, which will turn out to be related to the analogous limit for Kerr-Newman-like black holes in analogy to the minimal case, see section \ref{subsec:black holes no twist}.

\subsubsection*{The full partition function}

Let us now consider the previous results in the context of part II of the general conjecture. We gave a number of arguments that place the holographic round S$^3$ partition function at the NS limit of the gravitational Nekrasov building block \eqref{eq:Nek}, and naturally expect this to hold for the full quantum gravity version. We should however caution the reader that this gravitational statement should not be directly compared to the analogous limit in the refined topological string partition function used to calculate the holographically dual answer in \cite{Marino:2011eh,Hatsuda:2013oxa,Grassi:2014zfa,Hatsuda:2016uqa}. As far as we are aware, the reinterpretation of the field theory localization answer in terms of the topological string on {\it local} (i.e.\ {\it non-compact}) Calabi-Yau manifold is not understood systematically and should be considered a mathematical accident. 

Based on the on-shell action results, we were able to define the gravitational $\cF$-extremization above, which amounts to simply extremizing the result without considering the Coulomb branch parameters as being conjugate to any conserved charge (unlike the black holes cases). This is of course in perfect agreement with the fact that the symmetry pure AdS$_4$ does not allow for non-vanishing conserved charges. This however leads to a potential confusion when defining the gravitational partition function along the way proposed in part II of the conjecture. To show this explicitly, let us first formally evaluate what was defined as a {\it grand-canonical} partition function.
\be
\label{eq:GCsphere}
	Z(S^3_\text{sq.}, \chi^I, \omega) = Z^\text{q.g.}_\text{Nek} ((1+\omega)\, \chi^I, \omega)\ ,
\ee
under the constraint $g_I \chi^I = 1$, for the putative quantum gravitational Nekrasov partition function, $Z^\text{q.g.}_\text{Nek}$. We can also evaluate the {\it microcanonical} partition function
\be
	Z(S^3_\text{sq.}) =  \int \left( \prod_{I = 0}^{n_V} {\rm d} \chi^I \right)\, {\rm d} \omega\, \delta(g_I \chi^I - 1)\, Z(S^3_\text{sq.}, \chi^I, \omega)\ .
\ee
However, since the constrained integration actually does not mix the squashing parameter $\omega$ with the Coulomb branch parameters, it is also perfectly acceptable to define
\be
	Z(S^3_\text{sq.}, \omega) =  \int \left( \prod_{I = 0}^{n_V} {\rm d} \chi^I \right)\, \delta(g_I \chi^I - 1)\, Z(S^3_\text{sq.}, \chi^I, \omega)\ ,
\ee
as well as
\be
	Z(S^3_\text{sq.}, \chi^I) =  \int {\rm d} \omega\, Z(S^3_\text{sq.}, \chi^I, \omega)\ ,
\ee
under the constraint $g_I \chi^I = 1$. Since the saddle-point evaluation brings back the same leading contribution related to the on-shell action of the special solutions, we observe that at leading order
\be
\label{eq:Airycandidates}
	Z(S^3_\text{sq.}, \chi^I \Big|_\text{crit.}, \omega = 1) \approx Z(S^3_\text{sq.}, \chi^I \Big|_\text{crit.}) \approx Z(S^3_\text{sq.}, \omega=1) \approx  Z(S^3_\text{sq.}) \, ,
\ee
as well as
\be
\label{eq:squashedAirycandidates}
	Z(S^3_\text{sq.}, \chi^I \Big|_\text{crit.}, \omega)\approx Z(S^3_\text{sq.}, \omega) \, ,
\ee
and even
\be
\label{eq:CoulombAirycandidates}
	Z(S^3_\text{sq.}, \chi^I , \omega = 1) \approx  Z(S^3_\text{sq.}, \chi^I)\ .
\ee
The fact that the Laplace transform is trivial therefore presents us with a multiplicity of different partition functions that reproduce the same leading answer. Fortunately, the arising confusion of which one to pick in the correct holographic description is unique to the S$^3$ example and is non-existent for the black hole backgrounds (due to the non-trivial Laplace transform in those cases). 

Let us state clearly the arising conundrum: we have four equally viable candidates in \eqref{eq:Airycandidates} for the holographic dual of the Airy function \cite{Drukker:2011zy,Fuji:2011km,Marino:2011eh}, two candidates in \eqref{eq:squashedAirycandidates} for the holographic dual of the Airy function with squashing \cite{Hatsuda:2016uqa}, and another two candidates for the dual of the Airy function with general R-charge assignments \cite{Nosaka:2015iiw}. Note that the most general case including both squashing and general R-charge assignments, which can be uniquely identified with the grand-canonical partition function \eqref{eq:GCsphere} above, has not yet been fully understood field theoretically. There appear to exist interesting relations between the different limits of the general partition function, discovered recently in \cite{Chester:2021gdw}, which strongly suggests that the correct holographic identification makes use of the left-most object in each of the formulae \eqref{eq:Airycandidates} - \eqref{eq:CoulombAirycandidates}. Once the correct identification is clearly established, we would be able to use the impressive results established for ABJM theory in the cited references that would give precise predictions for the quantum corrections in $Z^\text{UV}_\text{Nek}$, including non-perturbative ones. Note that there are also many results on the three-sphere partition function for other holographic examples, see e.g.\ \cite{Jafferis:2011zi,Mezei:2013gqa,Fluder:2015eoa,Hosseini:2016ume} and references thereof, which could also provide interesting checks.

Finally, let us note that a related approach to the holographic dual of the Airy function was proposed in \cite{Dabholkar:2014wpa} based on supergravity localization and the two-dimensional theory \eqref{eq:2dSTUprepot}. Although there are many similarities in the final partition function expression, the idea there is that the Coulomb branch parameters are actually conjugate to the ABJM parameters $N$ and $k$. However, we seem to already find a disagreement with the way $\cF$-extremization works in this proposal, and therefore the present conjecture diverges considerably.

\subsection{Static/rotating black holes with a twist}
\label{subsec:black holes twist}
We now consider the static \cite{Cacciatori:2009iz,Katmadas:2014faa,Halmagyi:2014qza} and rotating \cite{Hristov:2018spe} BPS black holes with a twist in AdS$_4$. Supersymmetry is preserved by the so-called twisting condition, which translates into the fact that there is always a fixed non-zero magnetic charge carried by the R-symmetry gauge field. The resulting spacetime is therefore only asymptotically locally AdS$_4$, and due to the non-vanishing magnetic charge is sometimes called {\it magnetic} AdS$_4$ \cite{Hristov:2011ye,Hristov:2011qr}. Due to the twisting condition, one finds that the group of rotations commutes with the fermionic symmetries, and therefore in the spherical case the asymptotic group of symmetries corresponds to $\U(1|1) \times SO(3)$ \cite{Hristov:2013spa}. It follows that we can break the rotational symmetry without any further supersymmetry breaking, which allows the existence of rotating spherical solutions with axial symmetry or the change of horizon topology to an arbitrary Riemann surface $\Sigma_\mathfrak{g}$ of genus $\mathfrak{g}$ \cite{Caldarelli:1998hg}. Recently also the Coulomb branch solutions pertaining to the static black holes were constructed in Euclidean supergravity \cite{Bobev:2020pjk}. These solutions point strongly to the existence of the corresponding Coulomb branch solutions for the general rotating solutions as well, in agreement to the claim in part I of the conjecture. Due to the requirement that the background has fixed points under the canonical isometry, here we mainly need to consider the spherical rotating case, which was studied in detail at two derivatives in \cite{Hosseini:2019iad}. However, since it allows a smooth limit to the static case with a fixed two-submanifold S$^2$, which is turn closely related to the arbitrary genus case $\Sigma_\mathfrak{g}$, we will be able to describe all the aforementioned backgrounds.

\subsubsection*{Two derivative solutions}
As for the holographic sphere case, we are going to focus on the STU model,\footnote{Note that in \cite{Hosseini:2019iad} the STU model was defined with an opposite sign for the prepotential. This leads to a simple change in the overall sign in the building blocks \eqref{eq:conj1}, but more importantly does not change the definition of the quartic invariant $I_4$ (see also \cite{Hristov:2018spe}) used for performing the explicit calculations.} defined by
\be
	F_{2 \partial} = -2 \i\, \sqrt{X^0 X^1 X^2 X^3}\ , \quad g_0 = g_1 = g_2 = g_3 = 1\ , 
\ee
setting the AdS$_4$ length scale to $L = 1/\sqrt{2}$. The most general rotating black holes with magnetic AdS$_4$ asymptotics and spherical horizon were discovered in \cite{Hristov:2018spe}, where a flow from the 1/2 BPS AdS$_2 \times_w$S$^2$ near-horizon region to the asymptotic region was explicitly constructed. In fact we can focus purely at the near-horizon region, preserving the symmetry algebra $\SU(1,1|1) \times \U(1)$ (and additional flavor charges). Just like all other black hole spacetimes considered here, the fixed points of the canonical isometry sit precisely at the near-horizon, at the centre of the AdS$_2$ factor and the two poles of the sphere. The supersymmetric solutions were found using the integrability conditions following from a general time-like Killing spinor, see \cite{Cacciatori:2008ek,Meessen:2012sr}, and therefore we cannot present explicitly the corresponding Killing spinors and their bilinears. Nevertheless we can circumstantially infer that both fixed points come with the same orientation, or chirality, fixed here to be positive by convention. In the static case one can see that the full S$^2$ at the centre of AdS$_2$ becomes a fixed submanifold under the canonical isometry, as verified by the explicit Killing spinors found in \cite{DallAgata:2010ejj,Hristov:2010ri}. This is why the static limit indeed coincides with the bolt limit where the two points get blown up to a full two-submanifold as discussed earlier. The spinors therefore become chiral everywhere on the sphere, which is only possible if they had the same chirality at an arbitrary value of $\omega$ as well,
\be
	s_\text{SP} = s_\text{NP} = 1\ .
\ee
The rest of the {\it A}-gluing rule, as presented in table \ref{tab:1}, is ultimately justified by its success, which was shown in detail in \cite{Hosseini:2019iad} in the two derivative case. The above information is all we need to be able to generalize the two derivative result of \cite{Hosseini:2019iad} to the infinite derivative gluing, as discussed in part I of the conjecture.

Based on the STU model, the {\it A}-gluing rule and the resulting explicit black hole solutions of \cite{Hristov:2018spe}, \cite{Hosseini:2019iad} showed that the entropy function, in the notation and conventions introduced here, is given by
\be
\label{eq:twodertwistedentropyfun}
	\cI (\chi^I, \omega, p^I, q_I, \cJ) = \frac{\i \pi}{2 \omega G_N^{(4)}} \left( F(\chi^I+\omega p^I) - F(\chi^I - \omega p^I) \right) - \frac{\i \pi}{G_N^{(4)}} (\chi^I q_I - \omega \cJ)\ ,
\ee
under the constraints $\sum_I \chi^I = 1$ and $\sum_I p^I = -1$. In particular, it was shown that the black hole entropy is precisely reproduced upon extremizing the above function, in agreement with \eqref{eq:entropy}. Additionally, the critical values of the gluing parameters $X^I_{(1), (2)}$ are directly proportional to the background values of the on-shell sections at the poles of the sphere,
\be
	X^I_\text{SP} (z) = \i\, X^I_{(1)} \Big|_\text{crit.}\ , \qquad  X^I_\text{NP} (z) = \i\, X^I_{(2)} \Big|_\text{crit.}\ ,
\ee
such that reading off the background values of the sections for the black hole solutions allows one to check more easily the extremization of \eqref{eq:twodertwistedentropyfun}. Note that, unlike the case of the holographic sphere, \eqref{eq:twodertwistedentropyfun} is non-linearly dependent (involving square roots) on a number of independent charges.\footnote{In order to not deviate from the main point, we are not going to present explicitly the constraint $\hat{\lambda} (p^I, q_I, \cJ) = 0$ resulting from \eqref{eq:entropy}.} It is rather miraculous that the {\it A}-gluing reproduces exactly the supersymmetric background, providing very strong evidence of its applicability.\footnote{For the reader skipping the asymptotically flat examples, the {\it A}-gluing rule was also successfully tested with the cubic prepotential and FI gaugings allowing Minkowski$_4$ asymptotics, see \cite{Hosseini:2019iad}.} Note also that the analogous rule can be applied to the holographic partition functions that factorize in terms of holomorphic blocks and their $\omega \rightarrow 0$ limit, the {\it Cardy blocks} \cite{Choi:2019dfu}. This approach not only reproduces holographically the static results of \cite{Benini:2015eyy,Benini:2016rke}, but in the next-to-leading order in small $\omega$ is in agreement with the full expression above.

It is important to notice that the static limit is smooth not only at the level of the explicit two derivative solutions \cite{Hristov:2018spe}, but also at the level of the entropy function. When taking the limit $\omega \rightarrow 0$ we find a cancellation of the leading term in \eqref{eq:twodertwistedentropyfun}, resulting in a final answer that is $\omega$-independent,
\be
	\cI (\chi^I, p^I, q_I) = \lim_{\omega \rightarrow 0} \cI (\chi^I, \omega, p^I, q_I, \cJ) = \frac{\i \pi}{G_N^{(4)}}\, (p^I F_I (\chi^I) - \chi^I q_I)\ ,
\ee
with $\sum_I \chi^I = 1$ and $\sum_I p^I = -1$, in agreement with the earlier results of \cite{Cacciatori:2009iz,DallAgata:2010ejj,Benini:2015eyy}. We can also derive the entropy function for the case when we replace the spherical horizon with an arbitrary $\Sigma_\mathfrak{g}$ Riemann surface, simply by an overall rescaling \cite{Benini:2016hjo,Benini:2016rke},
\be
	\cI (\chi^I, p^I, q_I, \mathfrak{g}) = \frac{\eta}{2}\, \cI (\chi^I, p^I, q_I) = \frac{\i \pi\, \eta}{2 G_N^{(4)}}\, (p^I F_I (\chi^I) - \chi^I q_I)\ ,
\ee
where $\eta = 2 |\mathfrak{g}-1|$ ($\eta = 1$ for $\mathfrak{g} = 1$), the electric and magnetic charges obey a different quantization condition, and the constraint becomes $\sum_I p^I = -\rho$ for the constant scalar curvature $\rho$ (we already used the symbol $\kappa$ for the gravitational coupling constant). Notice that the corresponding on-shell action,
\be
\label{eq:staticonshell}
	\cF (\chi^I, p^I, \mathfrak{g}) = - \frac{\i \pi\, \eta}{2 G_N^{(4)}}\, p^I F_I (\chi^I)\ ,
\ee
is reproducing the result obtained in \cite{Bobev:2020pjk} for all the static Coulomb branch solutions. We expect the on-shell action corresponding to the rotating Coulomb branch solutions to be described by the Legendre transform of \eqref{eq:twodertwistedentropyfun}.

\subsubsection*{Higher derivatives and the static/bolt limit}
The full higher derivative prediction for the on-shell action of the twisted rotating black holes follows from the {\it A}-gluing applied to \eqref{eq:conj1}, resulting in
\be
\label{eq:twistedonshellhd}
\begin{split}
	\cF(p^I, \chi^I, \omega) = \frac{4 \i \pi^2}{\omega}\, \Big( &  F(\kappa^{-1} (\chi^I - \omega p^I); (1-\omega)^2, (1+\omega^2)) \\
&-  F(\kappa^{-1} (\chi^I + \omega p^I); (1+\omega)^2, (1-\omega^2)) \Big)\ ,
\end{split}
\ee
under the same constraints as in the two derivative case. Unfortunately, due to the fact that the rotating background was found only recently and the corresponding Killing spinors are not known explicitly, there is no available off-shell analysis of the rotating black holes and therefore we cannot justify the general expression in any rigorous way. The good news is that things are under better control in the static limit, because the static near-horizon geometry was analyzed off-shell in \cite{Hristov:2016vbm} in the presence of the $\mathbb{W}$ invariant (see also section 8 of \cite{deWit:2011gk}), and in \cite{Bobev:2020egg,Bobev:2021oku} in the limit to minimal supergravity in the presence of both $\mathbb{W}$ and $\mathbb{T}$ invariants at four derivatives. It is therefore useful to first write down the higher derivative prediction for the static/bolt limit, which gives
\be
\label{eq:boltconjecture}
	\cF_\text{bolt} (\mathfrak{g}, p^I, \chi^I) = - 4 \i \pi^2\, \eta\, \Big(\kappa^{-1}\, p^I F_I(\kappa^{-1} \chi^I; 1, 1) +2 \rho\,  F_{A_\mathbb{W}} (\kappa^{-1} \chi^I; 1, 1) -2 \rho\, F_{A_\mathbb{T}} (\kappa^{-1} \chi^I; 1, 1) \Big)\ ,
\ee
under the constraint $g_I \chi^I = 1$, where we already inserted the prefactor of $\eta = 2 |\mathfrak{g}-1|$ ($\eta = 1$ for $\mathfrak{g} = 1$), generalizing the spherical case to an arbitrary $\Sigma_\mathfrak{g}$ in analogy to the two derivative case. This also means that the magnetic charges obey $g_I p^I = -\rho$ as discussed above. It is natural to conjecture that this is the general answer for the contribution of a single fixed two-submanifold, or bolt, in the case of trivial fibration $\mathfrak{p} = 0$.

When applied to the case of minimal supergravity with a four derivative prepotential as in \eqref{eq:RHCBprepot} with $g_0 = 2$, the formula above simplifies further to
\be
	\cF (\mathfrak{g}, p^I, \chi^I) = (1-\mathfrak{g} )\, \left( \frac{\pi}{2 G_N^{(4)}} - 16 \pi^2\, c_1 + 16 \pi^2\, c_2 \right)\ ,
\ee
which is again in precise agreement (see footnote \ref{ft:25}) with the corresponding results in \cite{Bobev:2020egg,Bobev:2020zov,Bobev:2021oku}.

Coming back to the general matter coupled result \eqref{eq:boltconjecture} in the static limit, we find that it is in full agreement with the entropy function proposed in \cite{Hristov:2016vbm} based on a direct evaluation of the Wald entropy in the presence of an arbitrary tower of $\mathbb{W}$ terms. Based on the full off-shell background discussed partially already in \cite{deWit:2011gk} and completed in \cite{Hristov:2016vbm}, we can actually confirm the expectations for both composite fields $A_{\mathbb{W}}$ and $A_{\mathbb{T}}$. Focusing back to the spherical case, the near-horizon metric and the non-vanishing components of the $T$-tensor and the field strengths can be conveniently parametrized by
\be
	{\rm d} s^2 = v_1\, {\rm d} s^2_{AdS_2} + v_2\, {\rm d} s^2_{S^2}\ , \quad T^-_{01} = \i\, T^-_{23} = - w\ , \quad F^I_{23} = \frac{p^I}{v_2}\ ,
\ee
which is in full agreement with the ansatz for the BPS black holes in ungauged supergravity, c.f.\ section \ref{subsec:osv2}. The resulting solution looks simpler upon choosing the gauge fixing condition \eqref{eq:partialgauge},
\be
	v_1 = \frac{e^{- \cK_{2 \partial} (z, \bar{z})}}{4}\ , \qquad w = 4 \i e^{\cK_{2 \partial} (z, \bar{z})/2}\ .
\ee
The $T$-tensor therefore becomes precisely equal to the choice we made in the asymptotically flat case in section \ref{subsec:osv2}, which retroactively justifies it (we claimed the gauge fixing in the ungauged case agrees with the general choice \eqref{subsec:osv2}). The rest of the background solution is however different, and after a number of straightforward simplifications following from the gauge fixing choice, we find
\be
	R = 2 (v_1^{-1} - v_2^{-1})\ , \qquad D = - \frac16\, (v_1^{-1} + 2 v_2^{-1})\ , 
\ee
and
\be
	g_I \cF^{I, -}_{23} = \frac14\, (v_1^{-1} + 2 v_2^{-1}) \ , \qquad g_I Y^I_{i j} = \i\, v_2^{-1} \varepsilon_{i k} \sigma^3{}^k{}_j \ .
\ee
Remarkably, after some additional cancellations, this set of background fields result in
\be
	A_\mathbb{W} = A_\mathbb{T} = e^{\cK_{2\partial} (z, \bar{z})}\ .
\ee
These values are again in agreement with the general expression in \eqref{eq:boltconjecture}, showing the same proportionality between the on-shell action evaluation and the background values of $A_{\mathbb{W},\mathbb{T}}$ as in sections \ref{subsec:osv2} and \ref{subsec:sphere}. Note that upon allowing a general Riemann surface $\Sigma_\mathfrak{g}$ with a constant curvature $\rho$, one needs to substitute $v_2^{-1}$ with $\rho\, v_2^{-1}$ above.

Finally, let us note that unlike the cases in sections \ref{subsec:osv2} and \ref{subsec:sphere} we are not aware of a general proof that possible $D$-term corrections vanish for the twisted black holes considered here. On the other hand, this is guaranteed holographically, since the black holes considered here correspond to the so called topologically twisted index, see below. It can be shown in full generality, see e.g.\ \cite{Zaffaroni:2019dhb}, that the index is a purely holomorphic quantity that translates into the entropy function being influenced only by chiral quantities, i.e.\ $F$-terms.

\subsubsection*{Quantum entropy function and holography}
According to part II of the conjecture, the quantum entropy function in this case reads
\be
\label{eq:twistedQEF}
	Z (p^I, q_I, \cJ) := \int \left( \prod_{I = 0}^{n_V} {\rm d} \chi^I \right)\, {\rm d} \omega\, \delta(g_I \chi^I - 1)\, e^{-\cF(p^I, \chi^I, \omega)- \frac{8 \i \pi^2}{\kappa^2} (\chi^I q_I - \omega \cJ)}\, Z^\text{UV} (p^I, \chi^I, \omega)\ ,
\ee
with the general on-shell action written in \eqref{eq:twistedonshellhd}, $g_I p^I = -1$, and $\hat{\lambda} (g_I, q_I, \cJ) = 0$ resulting implicitly from \eqref{eq:entropy} applied to \eqref{eq:twistedonshellhd}. The UV correction proposed here follows again from the {\it A}-gluing rule and the putative Nekrasov building block $Z^\text{UV}_\text{Nek}$. It is also natural to propose that the static limit of this expression can also be generalized to include the static black hole solutions with arbitrary horizon topologies,
\be
\label{eq:statictwistedQEF}
	Z (\mathfrak{g}, p^I, q_I) := \int \left( \prod_{I = 0}^{n_V} {\rm d} \chi^I \right)\,\delta(g_I \chi^I - 1)\, e^{-\cF(\mathfrak{g}, p^I, \chi^I)- \frac{8 \i \pi^2}{\kappa^2}\, \chi^I q_I}\, Z^\text{UV} (\mathfrak{g}, p^I, \chi^I)\ ,
\ee
under the same constraints as above. As already discussed in the introduction, the form of the static quantum entropy function proposed here formally matches the result obtained from the supergravity localization program, in this case obtained in \cite{Hristov:2018lod,Hristov:2019xku}.\footnote{Note that these references only considered a two derivative prepotential, but due to the usage of the off-shell formalism one can directly extend the general results by including the full higher derivative on-shell action $\cF (\mathfrak{g}, p^I, \chi^I)$ discussed here.} As discussed earlier, the Coulomb branch solutions corresponding to the static black holes in question were constructed in \cite{Bobev:2020pjk}. They preserve the same asymptotic AdS$_4$ symmetries as the Lorentzian black hole solutions, but do not possess an infinite throat near the horizon and therefore do not enjoy conformal symmetry enhancement in the infra-red, when evoking the standard holographic RG flow interpretation. We can therefore see very explicitly that the proposed finite-dimensional integration above is actually a completely {\it on-shell} formula, matching precisely the analogous field theoretic picture. The identification with the supergravity localization results in \cite{Hristov:2018lod,Hristov:2019xku}, based on an off-shell fluctuations around the near-horizon background, therefore poses the question of reconciling the two pictures. We leave this as an open problem, which needs to be addressed after more evidence accumulates for the validity of part II of the conjecture. Provided a clear identification exists, the results of \cite{Hristov:2019xku,Hristov:2021zai} about the one-loop correction give an explicit example of what the general Nekrasov partition function $Z^\text{UV}_\text{Nek}$ should include, namely the loop corrections arising both from the massless four-dimensional effective action and from the full infinite set of massive KK modes. This is precisely in line with the discussion below Eq.\ \eqref{eq:qgNek}.

Let us also briefly comment on the holographically dual picture, which has been a subject of considerable interest initiated by the successful microscopic counting of the static black holes in maximal supergravity \cite{Benini:2015eyy,Benini:2016rke}. The relevant field theoretical quantity is the so-called topologically twisted index, which in the spherical case (including refinement by angular momentum), was defined in \cite{Benini:2015noa} as
\be
\label{eq:46}
	\exp \left( I_\text{RTTI} (\mathfrak{n}^I, \Delta^I, \varepsilon) \right) = {\rm Tr} (-1)^F\, e^{-\beta H} e^{\i\, \Delta^I J_I} e^{\i\, \varepsilon L_3}\ , 
\ee
where $J_I$ are the $U(1)_f$ flavor symmetry generators, and $L_3$ the $U(1)$ generator of rotations around the symmetry axis of S$^2$. The parameters $\mathfrak{n}^I, \Delta^I$ and $\varepsilon$, obeying the supersymmetric constraints $\sum_I \mathfrak{n}^I = 2$ and $\sum_I \Delta^I = 2 \pi$, are therefore the field theoretic analog of the black hole parameters $p^I, \chi^I$ and $\omega$, respectively. Note that the expression above is naturally defined in the grand-canonical ensemble, and therefore it is holographically dual precisely to the grand-canonical partition function
\be
\label{eq:47}
	Z (p^I, \chi^I, \omega) = e^{-\cF(p^I, \chi^I, \omega)}\, Z^\text{UV} (p^I, \chi^I, \omega)\ ,
\ee
which is just the inverse Laplace transform of \eqref{eq:twistedQEF}. A noteworthy progress was achieved in the static limit, which again admits a generalization to an arbitrary Riemann surface \cite{Benini:2016hjo},
\be
\label{eq:48}
	\exp \left( I_\text{TTI} (\mathfrak{g}, \mathfrak{n}^I, \Delta^I) \right) = {\rm Tr} (-1)^F\, e^{-\beta H} e^{\i\, \Delta^I J_I}\ , 
\ee
with the gravitational analog
\be
\label{eq:49}
	Z (\mathfrak{g}, p^I, \chi^I) =  e^{-\cF(\mathfrak{g}, p^I, \chi^I)}\, Z^\text{UV} (\mathfrak{g}, p^I, \chi^I)\ .
\ee
In the field theoretic large $N$ limit, corresponding to the two derivative approximation in supergravity, one can map precisely the holographically dual quantities  \cite{Hristov:2018lod,Hristov:2019xku}. Transforming the topologically twisted index to the microcanonical ensemble naturally results into the proposed index-extremization and consequent microscopic rederivation of the black hole entropy. This was initially achieved using the holographic duality between the gauged STU model and ABJM theory, but has since been generalized to other holographic examples \cite{Hosseini:2016ume,Guarino:2017pkw,Azzurli:2017kxo,Hosseini:2017fjo,Benini:2017oxt,Bobev:2018uxk,Gang:2019uay,Benini:2019dyp,Coccia:2020cku,Hosseini:2020mut,Hosseini:2020wag}. Due to the discovery of the Coulomb branch solutions \cite{Bobev:2020pjk}, again in the STU model, converting to the microcanonical ensemble no longer needed for a precise holographic match. It would therefore be interesting to explore fully the implications from the full holographic identification of \eqref{eq:46} with \eqref{eq:47} and in turn \eqref{eq:48} with \eqref{eq:49} beyond the leading order, extending the available subleading corrections discussed in \cite{Liu:2017vll,Jeon:2017aif,Liu:2017vbl,Liu:2018bac,PandoZayas:2019hdb,Gang:2019uay,Benini:2019dyp,Bobev:2021oku,Hristov:2021zai}. The identification of these grand-canonical partition functions would of course immediately imply a precise match of the respective microcanonical quantities, but is a slightly simpler task due to lack of additional integration over the chemical potentials.

\subsection{Kerr-Newman-like black holes without a twist}
\label{subsec:black holes no twist}
Let us now consider the other supersymmetric branch of asymptotically AdS$_4$ black hole solutions - the Kerr-Newman-like solutions with general electric and magnetic charges and rotation, found recently in \cite{Hristov:2019mqp}. These solutions generalize the previously known supersymmetric limit of the Kerr-Newman-AdS$_4$ background in minimal supergravity (see \cite{Kostelecky:1995ei,Caldarelli:1998hg}), as well as the electrically charged black holes in \cite{Cvetic:2005zi,Chow:2013gba} in the $X^0 X^1$ model (a truncation of the full STU model). These solutions are characterized by the vanishing R-symmetry background magnetic charge, thus placing them in a topologically distinct sector from the twisted black holes described in the previous subsection, see \cite{Hristov:2013spa}. The asymptotic spacetime, in the absense of flavor magnetic charges, is global AdS$_4$ with the supersymmetry algebra $\OSp(2|4)$ (in $\cN=2$ supergravity), such that the holographically dual quantity is the superconformal index \cite{Bhattacharya:2008zy,Kim:2009wb,Imamura:2011su}. In the presence of the extra flavor magnetic charges, leading to a deformation of the asymptotic symmetries, one should instead consider the generalized superconformal index, \cite{Kapustin:2011jm}. The supersymmetric black holes in this case do not admit a static limit and therefore the near-horizon geometry is necessarily a non-trivial fibration of AdS$_2$ over S$^2$ preserving $SU(1,1|1)$ symmetry. Just as in the other black hole examples the fixed point of the canonical isometry are situated at the near-horizon geometry, the center of AdS$_2$ and the two poles of the sphere. The corresponding entropy function was understood in \cite{Choi:2018fdc} and later in \cite{Cassani:2019mms} in the absence of flavor magnetic charges, and generalized in \cite{Hosseini:2019iad} based on the two derivative gravitational blocks. In this case there is no knowledge of any potential Coulomb branch solutions and we will focus purely on the Lorentzian black holes.

\subsubsection*{Two derivative solutions}
We are again going to consider explicitly the STU model, precisely as defined in \eqref{eq:2dSTUprepot}, and briefly summarize the two derivative results of \cite{Hosseini:2019iad}. Unfortunately all the supersymmetric solutions were constructed via the integrability conditions in \cite{Cacciatori:2008ek,Meessen:2012sr} and therefore we have no access to the explicit Killing spinors in this case. The {\it id}-gluing rule was therefore devised on the basis of the gluing of the superconformal index in terms of holomorphic blocks in \cite{Beem:2012mb} and shown to reproduce correctly the black hole entropy. Concretely, \cite{Hosseini:2019iad} considered the entropy function
\be
\label{eq:twodertwistedentropyfunKN}
	\cI (\chi^I, \omega, p^I, q_I, \cJ) = - \frac{\i \pi}{2 \omega G_N^{(4)}} \left( F(\chi^I+\omega p^I) +F(\chi^I - \omega p^I) \right) - \frac{\i \pi}{G_N^{(4)}} (\chi^I q_I - \omega \cJ)\ ,
\ee
under the constraints $\sum_I \chi^I = (1+\omega)$ and $\sum_I p^I = 0$. Specialized to the STU model, the Kerr-Newman-like black hole entropy of \cite{Hristov:2019mqp} is reproduced upon extremizing the above function, in agreement with \eqref{eq:entropy}. The critical values of the gluing parameters $X^I_{(1), (2)}$ are related to the background values of the on-shell sections at the poles of the sphere,
\be
	(X^I_\text{SP} (z))^* = -\i\, X^I_{(1)} \Big|_\text{crit.}\ , \qquad  X^I_\text{NP} (z) = -\i\, X^I_{(2)} \Big|_\text{crit.}\ ,
\ee
similarly to the relations following from the {\it max}-gluing in \eqref{eq:identificationosv}. The explicit expression for the entropy function is non-linearly dependent on a number of independent charges,\footnote{Again, we are not going to present explicitly the constraint $\hat{\lambda} (p^I, q_I, \cJ) = 0$ resulting from \eqref{eq:entropy}.} and the fact that its extremization reproduces exactly the black hole entropy is a strong indication of the correct gluing rule. Note however that at two derivatives we cannot fix independetly the signs of $s$ and $\omega$ at the second fixed point in \eqref{eq:conj1}, and we are going to justify the present choice when considering higher derivatives.

We can also consider the Cardy limit, $\omega \rightarrow 0$, of the general expression presented above. It is easy to see that in this limit the magnetic charges actually drop out and we find
\be
	\cI (\chi^I, \omega, q_I, \cJ) = \lim_{\omega \rightarrow 0} \cI (\chi^I, \omega, p^I, q_I, \cJ) = -\frac{\i \pi}{\omega G_N^{(4)}}\, F(\chi^I) - \frac{\i \pi}{G_N^{(4)}} (\chi^I q_I - \omega \cJ)\ .
\ee
Notice that at two derivatives this coincides with the limit of vanishing flavor magnetic charges $p^I = 0$. In this case the entropy function is in agreement with the results in \cite{Choi:2018fdc} and \cite{Cassani:2019mms}. The Cardy and the related vanishing $p^I$/minimal supergravity limits were also explored holographically in \cite{Choi:2019zpz,Bobev:2019zmz,Nian:2019pxj,Benini:2019dyp}. A more general approach, based on \cite{Hosseini:2019iad}, of decomposing the holographic partition functions in {\it Cardy blocks} was undertaken in \cite{Choi:2019dfu}, which not only reproduces the Cardy limit above, but in the limit of small $\omega$ is actually in agreement with the full {\it id}-rule leading to \eqref{eq:twodertwistedentropyfunKN}.

\subsection*{Higher derivatives, the Cardy limit and the partition function}
The general Kerr-Newman-like black holes in AdS$_4$ were discovered only recently and they have not yet been understood in the off-shell formalism. The only results reported in the literature were already presented in the minimal/universal case in section \ref{subsec:universal}, which luckily is enough to fully fix the ambiguity mentioned above about the signs of $s$ and $\omega$ at the second fixed point. Let us first write down the full on-shell that follows from the {\it id}-gluing applied to \eqref{eq:conj1},
\bea
	\label{eq:knonshellhd}
\begin{split}
	\cF(\text{KN}, p^I, \chi^I, \omega) = \frac{4 \i \pi^2}{\omega}\, \Big( & F(\kappa^{-1} (\chi^I + \omega p^I); (1-\omega)^2, (1+\omega^2)) \\
&+  F(\kappa^{-1} (\chi^I - \omega p^I); (1-\omega)^2, (1+\omega^2)) \Big)\ ,
\end{split}
\eea
under the same constraints as in the two derivative case, $\sum_I \chi^I = (1+\omega)$ and $\sum_I p^I = 0$. In the minimal limit there are no flavor magnetic charges and this expression reduces automatically to \eqref{eq:mingluing} (with $g_0 = 2$ as in section \ref{subsec:universal}) that was verified explicitly in the four derivative case \cite{Bobev:2020egg,Bobev:2020zov,Bobev:2021oku}. This uniquely fixes 
\be
	s_{(2)} = +\ , \qquad \omega_{(2)} = \omega\ ,
\ee
for the {\it id}-gluing that was presented in table \ref{tab:1}. 

In the absence of other results for comparison, we finish the discussion with a couple of observations. First, the higher derivative on-shell action simplifies considerably in the Cardy limit,
\be
	\label{eq:Cardyknonshellhd}
	\cF_\text{Cardy}(\text{KN}, \chi^I, \omega) = \lim_{\omega \rightarrow 0} \cF(\text{KN}, p^I, \chi^I, \omega)  = \frac{8 \i \pi^2}{\omega}\, F (\kappa^{-1}\, \chi^I; 1, 1)\ ,
\ee
under the constraints $\sum_I \chi^I = 1$ and $\sum_I p^I = 0$. The latter constraint becomes superfluous since magnetic charges are invisible in the Cardy limit. In this case we find an interesting relation between Cardy limits of the holographic squashed sphere, \eqref{eq:cardysquashedsphere}, and the Kerr-Newman-like black holes,
\be
	\cF_\text{Cardy}(\text{KN}, \chi^I, \omega) = 2\, \cF_\text{Cardy} (S^3, \chi^I, \omega)\ ,
\ee
which generalizes the same relation discovered in the minimal/universal limit in \cite{Bobev:2020egg,Bobev:2020zov,Bobev:2021oku}, see section \ref{subsec:universal}.

Second, based on part II of the conjecture, we can now give a proposal for the grand-canonical and microcanonical partition functions,
\be
\label{eq:58}
	Z (\text{KN}, p^I, \chi^I, \omega) = e^{-\cF(\text{KN}, p^I, \chi^I, \omega)}\, Z^\text{UV} (\text{KN}, p^I, \chi^I, \omega)\ ,
\ee
with $Z^\text{UV}$ following from the same {\it id}-gluing of the proposed UV-completion of the gravitational Nekrasov partition function, and
\be
	Z(\text{KN}, p^I, q_I, \cJ) = \int \left( \prod_{I = 0}^{n_V} {\rm d} \chi^I \right)\, {\rm d} \omega\, \delta(g_I \chi^I - (1+\omega))\, e^{- \frac{8 \i \pi^2}{\kappa^2} (\chi^I q_I - \omega \cJ)}\, Z (\text{KN}, p^I, \chi^I, \omega)\ ,
\ee
with the constraints $g_I p^I = 0$ and $\hat{\lambda}^{KN} (g_I, q_I, \cJ) = 0$ resulting implicitly from \eqref{eq:entropy} applied to \eqref{eq:knonshellhd}. The latter quantity is the quantum entropy function for the Kerr-Newman-like black holes in AdS$_4$. In the holographically dual three-dimensional superconformal field theory one can define the generalized superconformal index, \cite{Imamura:2011su,Kapustin:2011jm}
\be
\label{eq:60}
	\exp \left( I_\text{SCI} (\mathfrak{n}^I, \Delta^I, \varepsilon) \right) = {\rm Tr} (-1)^F\, e^{-\beta H} e^{\i\, \Delta^I J_I} e^{\i\, \varepsilon L_3}\ , 
\ee
which in analogy to the topologically twisted index is a holomorphic function of the chemical potentials $\Delta^I$. We can therefore argue again that $D$-terms cannot contribute to the supergravity result due to the expected holographic equality between \eqref{eq:58} and \eqref{eq:60}.

\section*{Acknowledgements}
I am grateful to Seyed Morteza Hosseini, Stefanos Katmadas, Ivano Lodato, Valentin Reys and Alberto Zaffaroni for valuable comments on the draft, and for teaching me many of the present subjects in first place. I am supported in part by the Bulgarian NSF grants DN08/3, N28/5, and KP-06-N 38/11.

\bibliographystyle{JHEP}
\bibliography{Nekrasov.bib}

\end{document}